\documentclass[11pt]{article}
\usepackage{bbm}

\usepackage{epsfig,endnotes}
\usepackage{multirow}
\usepackage{subfig}
\usepackage{graphicx}
\usepackage{grffile}

\newcounter{subcopyrightbox@save}
\usepackage[font=bf]{caption}
\usepackage{color, url}
\usepackage{xspace} 

\usepackage{amsmath}

\usepackage{mathrsfs}
\usepackage{amssymb}
\usepackage{amsmath}
\usepackage{amsthm}
\usepackage{epstopdf}
\usepackage{balance}
\usepackage{bm}
\usepackage{thm-restate,thmtools}

\theoremstyle{definition}
\newtheorem{definition}{Definition}[section]

\usepackage{mathtools}

\usepackage{geometry}
\usepackage{changepage}

\usepackage{booktabs}
\usepackage[hidelinks]{hyperref}
\usepackage{cleveref}
\usepackage{footmisc}
\usepackage{makecell}

\newcommand{\caseI}{\text{Case I}}
\newcommand{\caseII}{\text{Case II}}
\newcommand{\caseIII}{\text{Case III}}
\newcommand{\LP}{\text{Case III-LP}}
\newcommand{\FT}{\text{Case III-FT}}
\DeclareGraphicsExtensions{%
    .png,.PNG,%
    .pdf,.PDF,%
    .jpg,.mps,.jpeg,.jbig2,.jb2,.JPG,.JPEG,.JBIG2,.JB2}

\usepackage{tikz}
\usepackage{amsmath}
\usepackage{epsfig,endnotes}
\usepackage{multirow}

\usepackage{graphicx}
\usepackage{subfig}
\usepackage{grffile}

\usepackage{caption}
\usepackage{color, url}
\usepackage{xspace} 

\usepackage{mathrsfs}

\usepackage{amsmath}
\usepackage{epstopdf}
\usepackage{balance}
\usepackage{bm}
\usepackage{makecell}
\usepackage{cleveref}

\usepackage{algorithm}
\usepackage{algorithmic}
\usepackage{amsthm}
\usepackage{amssymb}

\allowdisplaybreaks
\DeclareMathAlphabet{\mathcal}{OMS}{cmsy}{m}{n}
\newcommand{\myparatight}[1]{\smallskip\noindent{\bf {#1}:}~}
\usepackage{filecontents}

\newcommand{\RN}[1]{%
  \textup{\uppercase\expandafter{\romannumeral#1}}%
}

\DeclareMathOperator*{\argmax}{arg\,max}

\AtBeginDocument{%
  \providecommand\BibTeX{{%
    \normalfont B\kern-0.5em{\scshape i\kern-0.25em b}\kern-0.8em\TeX}}}

\begin{document}
\begin{center}
{\Large{\bf{Pre-trained Encoders in Self-Supervised Learning Improve Secure and Privacy-preserving Supervised Learning}}}

\begin{tabular}{cccc}
&&&\\
Hongbin Liu$^*$ &&& Wenjie Qu$^*$ \\
Duke University &&& Huazhong University of Science  \\
hongbin.liu@duke.edu&&& and Technology \\
&&& wen\_jie\_qu@outlook.com \\
\\
Jinyuan Jia &&& Neil Zhenqiang Gong \\
University of Illinois Urbana-Champaign &&& Duke University\\
jinyuan@illinois.edu &&& neil.gong@duke.edu\\
\end{tabular}
\end{center}
\def\thefootnote{*}\footnotetext{Equal contribution. Wenjie Qu performed this research when he was a remote intern in Gong’s group.}
\begin{abstract}

Classifiers in supervised learning have various security and privacy issues, e.g., 1) \emph{data poisoning attacks}, \emph{backdoor attacks}, and \emph{adversarial examples}  on the security side as well as 2) \emph{inference attacks}  and \emph{the right to be forgotten} for the training data on the privacy side. Various secure and privacy-preserving supervised learning algorithms with formal guarantees have been proposed to address these issues. However, they suffer from various limitations such as accuracy loss, small certified security guarantees, and/or inefficiency. Self-supervised learning is an emerging technique to pre-train encoders using unlabeled data. Given a pre-trained encoder as a feature extractor, supervised learning can train a simple yet accurate classifier using a small amount of labeled training data. In this work, we perform the first systematic, principled measurement study to understand whether and when a pre-trained encoder can address the limitations of secure or privacy-preserving supervised learning algorithms. Our key findings are that  a pre-trained encoder substantially improves 1) both  accuracy under no attacks and certified security guarantees against data poisoning and backdoor attacks of state-of-the-art secure learning algorithms (i.e., bagging and KNN), 2)  certified security guarantees of randomized smoothing against adversarial examples without sacrificing its  accuracy under no attacks, 3)   accuracy of  differentially private classifiers, and 4)  accuracy and/or efficiency of exact machine unlearning. 

\end{abstract}
\section{Introduction}
\label{sec:intro}
Supervised learning is widely used in many security-critical and privacy-sensitive applications such as autonomous driving, cybersecurity, and healthcare. In a nutshell, supervised learning trains a classifier on labeled training data, which is then used to classify testing inputs. However, many studies showed that supervised learning has various security and privacy issues. For security, an attacker can leverage \emph{data poisoning attacks}~\cite{nelson2008exploiting,biggio2012poisoning},  \emph{backdoor attacks}~\cite{gu2017badnets,chen2017targeted,liu2017trojaning}, or  \emph{adversarial examples}~\cite{szegedy2013intriguing,carlini2017towards}  to make a classifier predict incorrect labels as an attacker desires. For privacy,  an attacker can leverage various inference attacks such as \emph{model inversion attacks}~\cite{fredrikson2015model}, \emph{membership inference attacks}~\cite{shokri2017membership}, and \emph{property inference attacks}~\cite{ateniese2015hacking,ganju2018property} to infer sensitive information about a classifier's training data. Moreover, various regulations and laws such as the European Union's General Data Protection Regulation (GDPR)~\cite{voigt2017eu} confer \emph{the right to be forgotten} to users, where a user's data should be deleted from the training data and its effect on a classifier should be removed upon the request from the user.

To address those security and privacy issues, various defenses have been proposed. Roughly speaking, those defenses can be divided into \emph{empirical defenses} and \emph{certified defenses}. Empirical defenses cannot provide  formal security or privacy guarantees, and thus they are often broken by advanced, adaptive attacks. Therefore, we focus on certified defenses in this work, which  provide formal security or privacy guarantees in adversarial settings.

\myparatight{Secure supervised learning algorithms and their limitations}
A certified defense against data poisoning and backdoor attacks is essentially a secure supervised learning algorithm, which can build a classifier on  a  training dataset whose predicted label for a testing input  is provably unaffected by attacks, once the number of added, deleted, and/or modified training examples is no larger than a threshold (called \emph{certified poisoning size}). In this work, we focus on \emph{bagging}~\cite{jia2021intrinsic} and \emph{K-nearest neighbors (KNN)}~\cite{jia2022certified}, which are two state-of-the-art secure supervised learning algorithms. 
A certified defense against adversarial examples can build a classifier that provably predicts the same label for a testing input, once the magnitude (e.g., measured by $\ell_2$-norm) of the adversarial perturbation is smaller than a threshold (called \emph{certified radius}). We focus on \emph{randomized smoothing}~\cite{cohen2019certified}, a state-of-the-art certified defense against adversarial examples that is applicable to arbitrary classifiers including large neural networks. 

Existing secure supervised learning algorithms suffer from two limitations: 1) they sacrifice the testing accuracy under no attacks, i.e., the learnt secure classifiers  are (much) less accurate than their non-secure counterparts when there are no attacks, and 2) their certified security guarantees are weak, i.e., they can only achieve  small certified poisoning sizes or certified radii. These limitations are the major obstacles for deploying them in practice.

\myparatight{Privacy-preserving supervised (un)learning algorithms and their limitations} To defend against inference attacks that compromise the training data privacy of a classifier, many studies~\cite{abadi2016deep,song2013stochastic,bassily2014private,chaudhuri2011differentially,kifer2012private,liu2020intrinsic} developed differentially private supervised learning algorithms to train the classifier with formal privacy guarantees. For instance, DP-SGD~\cite{abadi2016deep} is a state-of-the-art differentially private learning algorithm, which is applicable to deep neural networks. Exact machine unlearning aims to completely remove the effect of a user's data on a classifier, which certifiably achieves the right to be forgotten. \emph{Retraining from scratch} is the only exact machine unlearning method that is applicable to any learning algorithms. Roughly speaking, retraining from scratch first deletes a user's data from the training dataset and then trains a classifier from scratch on the remaining training data. 

The major limitation of existing differentially private learning algorithms is that a differentially private classifier has a (much) lower testing accuracy than the non-private counterpart. Retraining from scratch is either inefficient or the unlearnt classifier is inaccurate.

\myparatight{Self-supervised learning}Self-supervised learning~\cite{hadsell2006dimensionality,he2020momentum,chen2020simple,radford2021learning} is an emerging machine learning paradigm, which pre-trains an encoder using unlabeled data. The pre-trained encoder can be used as a feature extractor for supervised learning tasks. In particular, with high-quality feature vectors produced by the encoder, we can use supervised learning to train a simple yet accurate classifier with a small amount of labeled training data.

\myparatight{Our work} We perform the first systematic and principled measurement study to understand whether and when a pre-trained encoder in self-supervised learning can improve secure and privacy-preserving supervised (un)learning algorithms. We consider the pre-trained encoder can produce high-quality feature vectors. Moreover, the encoder is pre-trained on public data and thus does not incur privacy concerns for users. Our results show that the pre-trained encoder can address the limitations of a secure or privacy-preserving supervised (un)learning algorithm.

{\bf Measurement setup.} We adopt a state-of-the-art encoder pre-trained by OpenAI~\cite{radford2021learning} on 400 million (image, text) pairs collected from the public Internet. Moreover, we consider three benchmark datasets, i.e., STL10, CIFAR10, and Tiny-ImageNet, for supervised learning. To understand the benefits of a pre-trained encoder, we compare three cases, namely {\caseI}, {\caseII}, and {\caseIII}, on each dataset. In particular, we respectively train a linear classifier and a deep neural network from scratch in {\caseI} and {\caseII}  without using a pre-trained encoder. {\caseIII} considers a pre-trained encoder is available, and a common way to build a classifier is to concatenate the encoder and a fully connected layer with softmax activation function~\cite{bishop2006pattern}. Then, we train the classifier by either  just training the last fully connected layer (denoted as \emph{linear probing}) or training both the encoder and the last fully connected layer (denoted as \emph{fine-tuning}). Note that it is commonly believed that fine-tuning often achieves higher testing accuracy than linear probing in non-adversarial settings, which were confirmed by both prior work~\cite{kumar2021fine} and our work. As hyperparameters significantly influence the performance of a classifier~\cite{wang2018stealing}, we use grid search to find the best combination of the key hyperparameters of a secure or privacy-preserving algorithm in each case on each dataset. 

{\bf A pre-trained encoder improves secure supervised learning algorithms.} 
Via a systematic measurement study, we find that a pre-trained encoder substantially mitigates the limitations of secure supervised learning algorithms. Specifically, 
we find that a pre-trained encoder substantially improves both 1) the testing accuracy under no attacks of bagging and KNN and 2) their certified security guarantees against data poisoning and backdoor attacks. For instance, on STL10 dataset, the testing accuracies  of bagging for Case I, Case II, and Case III with linear probing are respectively  0.352, 0.559, 0.979, while the average certified poisoning sizes of the correctly classified testing examples are respectively 1.3, 2.8, and 68.8 in three cases. We also find that linear probing and fine-tuning in Case III achieve comparable testing accuracy and certified security guarantees for bagging, but linear probing is orders of magnitude more space and time efficient.  

Moreover, we find that a pre-trained encoder improves the certified security guarantees of randomized smoothing against adversarial examples without sacrificing its testing accuracy under no attacks. For example, on STL10, the average certified radii of the correctly classified testing examples are respectively 0.295, 0.625, and 1.482 in Case I, Case II, and Case III with fine-tuning, while the testing accuracies under no attacks are respectively 0.272, 0.610, and 0.612 in the three cases. Unlike bagging, we find that fine-tuning achieves much larger testing accuracy and  certified security guarantees  than linear probing for randomized smoothing when a pre-trained encoder is available. 

We note that state-of-the-art certified defense methods (i.e., bagging, KNN, and randomized smoothing) are based on ensemble. Their key limitation is that their base classifiers used to build ensemble classifiers are less accurate when trained on a small amount of subsampled training data or noisy training data. Our intuition is that pre-trained encoder can extract high-quality feature vectors and thus make their base classifiers more accurate. Note that more accurate base classifiers can lead to better security-utility tradeoff for those methods. 

{\bf A pre-trained encoder improves privacy-preserving supervised (un)learning algorithms.} We find that a pre-trained encoder also  substantially mitigates the limitations of privacy-preserving supervised (un)learning algorithms. Specifically, we find that a pre-trained encoder substantially improves the testing accuracy of a differentially private classifier. For instance, a differentially private classifier in {\caseIII} with linear probing  achieves a testing accuracy of 0.956 on STL10  under $(1.0,10^{-5})$-differential privacy. Under the same privacy budget, the testing accuracies of the differentially private classifiers in {\caseI} and {\caseII} are respectively 0.237 and 0.142. We also find that linear probing achieves much higher testing accuracy than fine-tuning to train a differentially private classifier when a pre-trained encoder is available. Our intuition is that with an encoder pre-trained on public, non-private data, we can train an accurate, yet privacy-preserving downstream linear classifier.

For exact machine unlearning, we find that retraining from scratch in {\caseIII} with linear probing is much more accurate and slightly more efficient than that in  {\caseI}, and is more accurate and orders of magnitude more efficient than that in {\caseII}. We also observe that linear probing is orders of magnitude more efficient than fine-tuning for retraining from scratch when a pre-trained encoder is available.

Our contributions can be summarized as follows:

\begin{itemize}
    \vspace{-1mm}
    \item We perform the first systematic and principled measurement study to understand whether and when a pre-trained encoder in self-supervised learning improves  secure or privacy-preserving supervised (un)learning algorithms.
    \vspace{-1mm}
    \item We find that a pre-trained encoder substantially improves the testing accuracy under no attack and the certified security guarantees of state-of-the-art provably secure supervised learning algorithms, i.e., bagging, KNN, and randomized smoothing.
    \vspace{-1mm}
    \item We find that a pre-trained encoder substantially improves the testing accuracy of a differentially private classifier, and the efficiency or the testing accuracy of exact machine unlearning. 
\end{itemize}  
\section{Background}
\label{sec:background}
\subsection{Self-Supervised Learning}

Given a \emph{pre-training dataset}, e.g., a set of unlabeled images or (image, text) pairs, 
self-supervised learning~\cite{hadsell2006dimensionality,he2020momentum,chen2020simple,radford2021learning} aims to  pre-train an image encoder and (optionally) a text encoder, which can be used as general-purpose feature extractors for various AI tasks. Many self-supervised learning algorithms have been developed in the past several years. Among them,  CLIP~\cite{radford2021learning} is the state-of-the-art one. Specifically, CLIP jointly pre-trains an image encoder and a text encoder using a set of (image, text) pairs. The image encoder and text encoder respectively output a feature vector for an image and a text. Given a mini-batch of $B$ (image, text) pairs, we use the image encoder to produce a feature vector for each of the $B$ images and the text encoder to produce a feature vector for each of the $B$ texts.  
CLIP aims to jointly optimize the image encoder and text encoder such that an image and a text in the same (image, text) pair have similar feature vectors, while an image and a text from different (image, text) pairs have dissimilar feature vectors. OpenAI uses CLIP to pre-train an image encoder and a text encoder on 400 million (image, text) pairs collected from the public Internet~\cite{radford2021learning}.

\subsection{Supervised Learning}
\label{sec-supervised-learning}
Suppose we have a labeled training dataset $\mathcal{D}= \{(\mathbf{x}_1,y_1), (\mathbf{x}_2,  y_2), \cdots, (\mathbf{x}_n,y_n)\}$, where $\mathbf{x}_i\in \mathbb{R}^d$ is the $i$-th training input, $y_i$ is the corresponding label, and $d$ is the input dimension. Supervised learning aims to train a classifier $f$ on the training dataset. In particular, the classifier $f$ is a function that maps an input $\mathbf{x}\in \mathbb{R}^d$ to a label among a set of $c$ labels $\{1,2,\cdots,c\}$.  We consider two scenarios depending on whether a pre-trained encoder is used or not.

\myparatight{Without a pre-trained encoder} 
Given the training dataset $\mathcal{D}$, supervised learning defines a loss function $\mathcal{L}$ for a classifier $f$: $\mathcal{L} = \frac{1}{n} \sum_{i=1}^{n}l(f(\mathbf{x}_i),y_i)$, where ${l}$ is the loss function, e.g., cross-entropy loss, for a single training example. Supervised learning trains the classifier $f$ via minimizing the loss function $\mathcal{L}$, e.g., using  stochastic gradient descent (SGD).

\myparatight{With a pre-trained encoder} When a pre-trained encoder is available, we can use it as a feature extractor when training a classifier. In general, given a pre-trained encoder, there are two popular ways~\cite{chen2020simple,chen2020big,he2020momentum,radford2021learning}, i.e., \emph{linear probing (LP)} and \emph{fine-tuning (FT)}, to train a classifier depending on whether the parameters of the pre-trained encoder are updated or not during training. In particular, LP freezes the parameters of the pre-trained encoder during training, while FT updates the parameters of the pre-trained encoder during training. Next, we  discuss them in more details.
\begin{itemize}
    \vspace{-1mm}
    \item {\bf Linear probing (LP):} LP appends a fully connected layer with the sofmax activation function to the pre-trained encoder, where the number of output neurons is the number of classes for the classification task. The composition of the pre-trained encoder and the fully connected layer is viewed as a  classifier $f$, which maps an input to a label. When training the classifier using supervised learning, LP freezes the parameters of the pre-trained encoder and only updates the parameters of the fully connected layer. 
    
    \item {\bf Fine-tuning (FT):} Like LP, FT also appends a fully connected layer with  the sofmax activation function to the pre-trained encoder. However, unlike LP, FT updates the parameters of both the pre-trained encoder and the fully connected layer when training the classifier using supervised learning. 
\end{itemize}

\section{Threat Model}
\myparatight{Classifiers in supervised learning} We consider that a classifier in supervised learning is vulnerable to various security and privacy attacks.  In particular, for security, a classifier is  vulnerable to \emph{data poisoning attacks}~\cite{nelson2008exploiting,biggio2012poisoning}, \emph{backdoor attacks}~\cite{gu2017badnets,chen2017targeted,liu2017trojaning}, or \emph{adversarial examples}~\cite{szegedy2013intriguing,carlini2017towards}. A data poisoning attack assumes that an attacker can  tamper with the labeled training dataset such that the learnt classifier makes incorrect predictions for indiscriminate clean testing inputs or attacker-chosen ones. In a backdoor attack, an attacker may also tamper with the training dataset but the backdoored classifier predicts an attacker-chosen label for any input embedded with a backdoor trigger. In an adversarial example, an attacker adds a carefully crafted, human-imperceptible perturbation to a testing input such that it is misclassified as the attacker desires. Various defenses have been proposed to defend against these attacks. We consider \emph{certified defenses} in this work, because they have formal security guarantees against these attacks.  

For privacy, given a black-box or white-box access to a classifier, an attacker can use various inference attacks such as \emph{model inversion attacks}~\cite{fredrikson2015model}, \emph{membership inference attacks}~\cite{shokri2017membership}, and \emph{property inference attacks}~\cite{ateniese2015hacking,ganju2018property}, to infer sensitive information about the training data of the classifier. Differential privacy is the de-facto standard for privacy protection, which has formal privacy guarantees. Therefore, we consider training a classifier with differential privacy to protect the  training data privacy. Moreover, a user may request deleting his/her data from the training dataset due to privacy concerns after a classifier has been trained. Upon such request, the classifier has to be unlearnt to remove the effect of the user's data, which is known as the right to be forgotten. We consider \emph{exact machine unlearning}, which has formal privacy guarantees at achieving the right to be forgotten.  

\myparatight{Pre-trained encoders}
We consider an encoder is pre-trained on public, non-sensitive data in non-adversarial settings. In other words, we assume 1) the pre-trained encoder produces high-quality feature vectors for inputs, and 2) the pre-training dataset does not have privacy concerns for users. For instance, the publicly available image encoder pre-trained by OpenAI on public (image, text) pairs collected from the Internet~\cite{radford2021learning}  is an example of such encoder. We note that pre-trained  encoders could also be vulnerable to data poisoning and backdoor attacks~\cite{jia2021badencoder,carlini2021poisoning_clip}, which may reduce the quality of the produced feature vectors. However, security/privacy of pre-trained encoders is beyond the scope of our work. In this work, we aim to understand whether and when a clean, high-quality pre-trained encoder can improve a secure or privacy-preserving  supervised learning  algorithm.   

\section{Measurement Setup}
\label{sec:methodology}
\subsection{Experimental Setup}
\label{sec:experimentalsetup}

\myparatight{Pre-trained encoder (i.e., CLIP~\cite{radford2021learning})}
We use the state-of-the-art image encoder (called CLIP) pre-trained by OpenAI~\cite{radford2021learning} in our evaluation. Specifically, this image encoder is pre-trained by OpenAI on 400 million public (image, text) pairs collected from the Internet. The neural network architecture of the image encoder is a vision transformer~\cite{dosovitskiy2020image} and the encoder outputs a 1024-dimension feature vector for an input. We downloaded the encoder from GitHub~\cite{clip_github}.

\myparatight{Datasets for training and testing classifiers} 
We use three benchmark datasets, i.e., STL10 \cite{coates2011analysis}, CIFAR10~\cite{krizhevsky2009learning}, and Tiny-ImageNet~\cite{tinyimagenet}, to train and test classifiers. 
Each dataset contains a pre-defined training dataset and testing dataset. Moreover, Tiny-ImageNet also contains a pre-defined validation dataset. For STL10 or CIFAR10, we evenly split its pre-defined testing dataset into two halves, which we respectively treat as validation examples and testing examples. Table~\ref{dataset_details} shows the statistics of the three datasets. For each  dataset, we use its training examples to train a  classifier, use its validation examples to search hyperparameters (e.g., learning rate, number of training epochs) used to train the  classifier, and use its testing examples to evaluate the  classifier. Unless otherwise mentioned, we show experimental results on STL10. 
\begin{table}[!t]
\caption{Statistics of  datasets.}
\vspace{-5mm}
\label{dataset_details}
\begin{center}
\begin{tabular}{|c|c|c|c|}
\hline
         & STL10 &  CIFAR10  & Tiny-ImageNet\\ \hline
\#Classes   &   10      &  10       & 200         \\ \hline
\#Training examples  &   5,000     &  50,000       & 100,000       \\ \hline
\#Validation examples &   4,000      &   5,000      &   10,000    \\ \hline
\#Testing examples &   4,000      &   5,000      &   10,000    \\ \hline
\end{tabular}
\vspace{-5mm}
\end{center}
\end{table}

\myparatight{Three cases to train classifiers} Our goal is to measure whether a pre-trained encoder  improves a secure or privacy-preserving supervised learning algorithm. Therefore, we consider the following three cases to train classifiers, where a pre-trained encoder is not used in the first two cases while is used in the third case. 
\begin{itemize}
    \item
    \myparatight{{\caseI}} The first case is to train a simple linear classifier using the training examples. In particular, we train a multinomial logistic regression classifier, which can be viewed as a simple neural network with a fully connected layer with softmax activation function. We note that we represent each input image as a vector. For instance, in CIFAR10, an image has size  $32 \times 32 \times 3$, which can be represented  as a vector with $3,072$ dimensions. 
    \item
    \myparatight{{\caseII}} The second case is to train a deep neural network classifier from scratch. We adopt ResNet-18~\cite{he2016deep} as the architecture of the deep neural network  classifier because it achieves state-of-the-art accuracy on the three datasets. 
    \item
    \myparatight{{\caseIII} ({\LP} and {\FT})} The third case is to train a classifier using linear probing or fine-tuning when a pre-trained encoder is available (please see Section~\ref{sec-supervised-learning} for details). In particular, we append a fully connected layer with softmax activation function (i.e., a multinomial logistic regression classifier) to the encoder. The composition of the encoder and the fully connected layer is a classifier. Then, in LP, we only train the fully connected layer using supervised learning, while we train both the encoder and the fully connected layer in FT. We denote these two scenarios as \emph{{\LP}} and \emph{{\FT}}, respectively. 
\end{itemize}

Case I and Case II represent two extreme cases of classifiers (i.e., simple linear classifier and complex deep neural network classifier) when a pre-trained encoder is not available. Training a linear classifier is more efficient but less accurate than training a deep neural network. Comparing Case III with the first two cases enables us to comprehensively measure the benefits of a pre-trained encoder when training a classifier.   

\myparatight{Hyperparameter search} The performance of a classifier highly depends on the hyperparameters used to train the classifier. The best hyperparameters of a classifier may be different for different datasets. Moreover, the best hyperparameters for the same dataset may be different for the  three cases. Since our goal is to compare the classifiers in the three cases, we aim to search for the best hyperparameters for each case. Specifically, based on the performance of a secure or privacy-preserving classifier on the validation examples, we perform grid search to find the best combination of the number of training epochs,  learning rate, and security/privacy-dependent hyperparameters for each case on each dataset. 

Specifically,  the metric to evaluate the performance of a classifier and the security/privacy-dependent hyperparameters (e.g., the number of nearest neighbors $K$ in $K$-nearest neighbor algorithm)  are different for different secure and privacy-preserving  algorithms, whose details we will discuss  when introducing our measurement results for each secure or privacy-preserving algorithm in Section~\ref{sec-poison-backdoor} to Section~\ref{sec-unlearn}. We consider every 5 training epochs up to 500 ones and learning rates $10^{-5}, 10^{-4}, 10^{-3}, 10^{-2}$, and $10^{-1}$ in our grid search. We note that mini-batch size is also a hyperparameter for training a classifier. However, we found that mini-batch size has a small impact once it is large enough. Therefore, we set mini-batch size to be 128 and do not perform grid search for it  to reduce the computation cost of hyperparameter search.   

\vspace{-5mm}
\subsection{Accuracy in Non-adversarial Setting}

\begin{table}[!t]
\caption{The testing accuracy of the three cases in non-adversarial setting.}
\vspace{-5mm}
\label{comparison-non-adversarial}
\begin{center}
\begin{tabular}{|c|c|c|c|}
\hline
         & STL10 &  CIFAR10  & Tiny-ImageNet\\ \hline
\caseI   &   0.353      &  0.438    &    0.069    \\ \hline
\caseII  &   0.755      &  0.953     &   0.516    \\ \hline
\LP & 0.985    & 0.954       &      0.751     \\ \hline
\FT &  0.970     & 0.973     &     0.785     \\ \hline
\end{tabular}
\end{center}
\vspace{-5mm}
\end{table}

Table~\ref{comparison-non-adversarial} shows the testing accuracy of the classifiers in the three cases  under non-adversarial settings. First, we observe that {\caseIII} achieves larger testing accuracy than {\caseI} and {\caseII}. The reason is that the pre-trained encoder extracts high-quality feature representations, which help improve the testing accuracy. Our observation is consistent with prior studies on self-supervised learning~\cite{radford2021learning,chen2020simple}. Moreover, {\caseII} achieves larger testing accuracy than {\caseI}. This is because a deep neural network in {\caseII} is more expressive than a linear classifier in {\caseI}. Second, {\FT} outperforms {\LP} in general. Specifically, the average testing accuracy of {\FT} and {\LP} across the three datasets in our experiments is 0.909 and 0.897, respectively. 
Our observation is also consistent with prior works~\cite{kumar2021fine,kornblith2019better,zhai2019large} on comparing {\FT} and {\LP} in non-adversarial settings, which also found that  {\FT} outperforms {\LP} with respect to average testing accuracy across datasets.  {\FT} achieves a smaller testing accuracy than {\LP} on STL10. We suspect the reason is that STL10 only contains 5,000 training examples and {\FT} is more likely to overfit the classifier to the training examples.

\section{Security against Data Poisoning and Backdoor Attacks} 
\label{sec-poison-backdoor}
\subsection{Related Work}
\myparatight{Data poisoning and backdoor attacks}
 Classifiers are vulnerable to data poisoning attacks~\cite{nelson2008exploiting,biggio2012poisoning,geiping2020witches,munoz2017towards,shafahi2018poison} and backdoor attacks~\cite{gu2017badnets,chen2017targeted,liu2017trojaning}. In data poisoning attacks, an attacker poisons the training data (e.g., adds, deletes, and/or modifies training examples) such that a classifier trained on the poisoned training data has attacker-desired  behaviors. Specifically, the poisoned classifier either achieves a low testing accuracy for indiscriminate clean testing inputs~\cite{biggio2012poisoning,munoz2017towards,shan2020fawkes,fowl2021adversarial} or predicts attacker-chosen labels for attacker-chosen clean testing inputs~\cite{shafahi2018poison, carlini2021poisoning, carlini2021poisoning_clip}. Like data poisoning attacks, some backdoor attacks also poison the training data. However, unlike data poisoning attacks, a backdoored classifier predicts an attacker-chosen label for any testing input embedded with a backdoor trigger and its predictions for clean testing inputs without the backdoor trigger are unaffected. For instance, BadNet~\cite{gu2017badnets}  embeds a backdoor trigger (e.g., a pixel pattern) to some training inputs and changes their labels to the attacker-chosen label. The classifier trained on the poisoned training data predicts the attacker-chosen label for any testing input embedded with the same backdoor trigger.

\myparatight{Certified defenses against data poisoning and backdoor attacks} Based on whether a defense has formal security guarantee against data poisoning and backdoor attacks, we can categorize  existing defenses into \emph{empirical defenses}~\cite{cretu2008casting,rubinstein2009antidote,barreno2010security,paudice2018label,paudice2018detection,suciu2018does,liu2018fine,peri2020deep,chen2021pois,tran2018spectral} and \emph{certified defenses}~\cite{jia2021intrinsic,jia2022certified,rosenfeld2020certified,jia2020certified,wang2021certified,levine2020deep}. In particular, empirical defenses do not have formal security guarantees and thus they are often broken by more advanced, adaptive attacks~\cite{koh2022stronger}. Therefore, we focus on certified defenses in this work. 

A certified defense is a secure learning algorithm $\mathcal{A}$ whose predicted label for a testing input is provably not affected by poisoning training examples once the number of them is no larger than a threshold, which is called \emph{certified poisoning size}. Specifically, we denote $\mathcal{A}(\mathcal{D}, \mathbf{x})$ as the label predicted for a testing input $\mathbf{x}$ by a classifier learnt by $\mathcal{A}$ on the training dataset $\mathcal{D}$. In data poisoning and backdoor attacks, an attacker can arbitrarily add, delete, and/or modify some training examples in $\mathcal{D}$. For simplicity, we use $\mathcal{D}'$ to denote the poisoned training dataset. We denote by $P(\mathcal{D},\mathcal{D}')$ the number of poisoning training examples in $\mathcal{D}'$, compared to $\mathcal{D}$. 
A certified defense guarantees the following:
\begin{align}
    \mathcal{A}(\mathcal{D},\mathbf{x}) = \mathcal{A}(\mathcal{D}^{\prime},\mathbf{x}), \forall \mathcal{D}' \text{ s.t. } P(\mathcal{D},\mathcal{D}') \leq r_{\mathbf{x}},
\end{align}
where $r_{\mathbf{x}}$ is the certified poisoning size for $\mathbf{x}$. In data poisoning attacks, $\mathbf{x}$ is a clean testing input. In backdoor attacks, $\mathbf{x}$ is a testing input with a backdoor trigger. The above guarantee means once a classifier trained on the clean training dataset correctly classifies the testing input with trigger (i.e., $\mathcal{A}(\mathcal{D},\mathbf{x})$ is the true label of $\mathbf{x}$), the backdoored classifier   trained on the poisoned training dataset still correctly classifies the testing input with trigger (i.e., $\mathcal{A}(\mathcal{D}^{\prime},\mathbf{x})$ is still correct) when the number of poisoning training examples is at most $r_{\mathbf{x}}$. 

Next, we discuss two state-of-the-art certified defenses, namely \emph{bagging}~\cite{jia2021intrinsic} and \emph{K-nearest neighbors}~\cite{jia2022certified}. We respectively discuss how they build a secure classifier and how they compute the certified poisoning size for a testing input. 

\emph{\bf 1) Bagging~\cite{breiman1996bagging,jia2021intrinsic}.}
Bagging~\cite{breiman1996bagging} is a popular ensemble learning algorithm. Suppose we create a random \emph{subsample} with $k$ training examples via sampling them from a training dataset $\mathcal{D}$ uniformly at random with replacement. We use an arbitrary learning algorithm (either deterministic or probabilistic) to train a classifier (called \emph{base classifier}) on it. Due to randomness in the subsample and (the learning algorithm), the label predicted by the base classifier for a testing input $\mathbf{x}$ is also random. We define \emph{label probability} $p_i$ ($i=1,2,\cdots c$) as the probability that the base classifier predicts label $i$ for $\mathbf{x}$. Given the label probability $p_i$'s, bagging builds an \emph{ensemble classifier} $F$ which predicts the label (denoted as $F(\mathcal{D},\mathbf{x})$) with the largest label probability for the testing input $\mathbf{x}$. Formally, we have: $F(\mathcal{D},\mathbf{x}) = \argmax_{i=1,2,\cdots,c}p_i.$

Jia et al.~\cite{jia2021intrinsic} showed that bagging has certified security guarantees and leveraged Monte-Carlo sampling to compute $F(\mathcal{D},\mathbf{x})$ and $r_{\mathbf{x}}$. In particular, they first create $N$ subsamples, each of which includes $k$ training examples sampled from $\mathcal{D}$ uniformly at random with replacement, then train a base classifier on each subsample, and finally count the number of base classifiers that predict label $i$ for $\mathbf{x}$, where $i=1,2,\cdots, c$.   The label with the largest count is treated as  $F(\mathcal{D},\mathbf{x})$, i.e., the predicted label for $\mathbf{x}$. Jia et al. showed that the certified poisoning size $r_{\mathbf{x}}$ is the solution to an optimization problem, which relies on a lower bound of the largest label probability  and an upper bound of the second largest label probability for $\mathbf{x}$. Given the count of each label, Jia et al.~\cite{jia2021intrinsic} used a Monte-Carlo method to estimate them, which are then used to solve the optimization problem to obtain the certified poisoning size $r_{\mathbf{x}}$. Due to the randomness in estimating the label-probability bounds,  the obtained $r_{\mathbf{x}}$ may be incorrect with a probability $\alpha$, where $\alpha$ is called \emph{confidence level} that can be set to be arbitrarily small. Note that $\alpha$ influences the estimated certified poisoning size $r_{\mathbf{x}}$ but not the predicted label $F(\mathcal{D},\mathbf{x})$. 

\emph{2) \bf $K$-nearest neighbors (KNN)~\cite{altman1992introduction,jia2022certified}.} K-nearest neighbors (KNN) is a classical non-parametric algorithm for classification. Jia et al.~\cite{jia2022certified} showed that KNN is intrinsically secure against data poisoning and backdoor attacks. Given a training dataset $\mathcal{D}$, a distance metric $d$ (e.g., $\ell_1$ distance in our experiments), and a testing input $\mathbf{x}$, KNN first calculates the distance between each training input and $\mathbf{x}$ with respect to the distance metric $d$, and then finds the $K$ training examples that have the smallest distances to $\mathbf{x}$.  Given the $K$ nearest neighbors, KNN takes a majority vote among their labels as  the predicted label for $\mathbf{x}$.

Jia et al. derived the certified security guarantee of KNN. Suppose  $a$ (or $b$) is the most  (or second most) frequent label among the $K$ nearest neighbors of $\mathbf{x}$. Moreover, we  use $K^a_{\mathbf{x}}$ (or $K^b_{\mathbf{x}}$) to  denote the number of nearest neighbors that have label $a$ (or $b$). Jia et al. showed that $r_{\mathbf{x}}$ can be computed as follows:
\begin{align}
\label{formal_cps_knn}
    r_{\mathbf{x}} = \left\lceil\frac{K^a_{\mathbf{x}}-K^b_{\mathbf{x}}+\mathbb{I}(a>b)}{2}\right\rceil-1,
\end{align}
where $\mathbb{I}$ is an indicator function whose value is $1$ if the condition $a>b$ is true and is $0$ otherwise.

\myparatight{Limitations of bagging and KNN} Bagging and KNN have two limitations: 1) they sacrifice testing accuracy under no attacks, i.e., the (ensemble) classifiers built by bagging or KNN are (much) less accurate than non-secure classifiers (e.g., deep neural network) when there are no attacks, and 2) their certified poisoning sizes are small, i.e., they can only tolerate a small number of poisoning training examples. These two limitations substantially hinder their deployment in practice. We aim to understand whether and when a pre-trained encoder can mitigate these limitations. 

\subsection{Experimental Setup}
\myparatight{Building classifiers using bagging or KNN} Suppose we have a training dataset $\mathcal{D}$. For bagging, we can respectively use {\caseI}, {\caseII}, and {\caseIII} to train $N$ base classifiers on $N$ subsamples created from $\mathcal{D}$. Given those base classifiers, we build an ensemble classifier for each case.

For KNN, {\caseI}, {\caseII}, and {\caseIII} have different meanings as KNN does not have training process. In particular,  in {\caseI}, we directly apply KNN to the raw training/testing images, i.e., the distance between a testing image and a training image is calculated based on their pixel values. To further improve KNN, Jia et al.~\cite{jia2022certified} proposed to first extract histogram-of-oriented-gradients (HOG)~\cite{dalal2005histograms} features for each training/testing image and then calculate distance between two images based on those features in KNN, which we use in our {\caseII}. In {\caseIII}, we use a pre-trained encoder to compute a feature vector for each training/testing image and then calculate distance between two images based on the feature vectors in KNN. 

\myparatight{Evaluation metrics} We use \emph{testing accuracy under no attacks (testing accuracy)} and \emph{average certified poisoning size (ACPS)} as the evaluation metrics. Testing accuracy is the fraction of clean testing inputs whose labels are correctly predicted by a classifier when there are no attacks. ACPS is the average certified poisoning size of the testing inputs that are correctly classified. Formally, given a testing dataset $\mathcal{D}_{te}$, we compute $\text{ACPS}$  as follows:
\begin{align}
  ACPS &=\frac{\sum_{(\mathbf{x}_{i},y_i) \in \mathcal{D}_{te}} \mathbb{I}(c_{\mathbf{x}_i}=y_{i}) \cdot r_{\mathbf{x}_i}}{\sum_{(\mathbf{x}_{i},y_i) \in \mathcal{D}_{te}} \mathbb{I}(c_{\mathbf{x}_i}=y_{i})} ,
\end{align}
where $c_{\mathbf{x}_i}$ and $r_{\mathbf{x}_i}$ respectively represent the predicted label and the certified poisoning size of the testing input $\mathbf{x}_{i}$, and $\mathbb{I}$ is an indicator function. For data poisoning attacks, the testing dataset $\mathcal{D}_{te}$ contains clean testing examples. For backdoor attacks, the testing dataset $\mathcal{D}_{te}$ contains backdoored testing examples obtained via embedding a backdoor trigger to each testing input. Note that the ACPS does not account for the clean data samples used to pre-train the pre-train the CLIP encoder in our \caseIII, since we consider the security of supervised learning, i.e., classifiers.

\myparatight{Hyperparameter search and parameter setting} 
We search the best hyperparameter setting for each of the three cases to achieve the largest ACPS. 
In bagging,  $k$ is the number of training examples used to train each base classifier. As we will show in our experimental results, ACPS first increases and then decreases as $k$ increases. Moreover, the $k$ that achieves the largest ACPS could also be different for the three cases. To fairly compare the three cases, we search for a wide range of $k$ when training base classifiers for each case. In particular, we consider the following values of $k$: 10, 30, 50, 100, 300, 500, and 1,000. We use grid search to find the combination of $k$, number of training epochs, and learning rate  that achieves the largest ACPS following the search process described in Section~\ref{sec:experimentalsetup}. We note that there are two other hyperparameters in testing phase of bagging, i.e., $N$ and $\alpha$, which respectively control the number of base classifiers and  confidence level of the outputted certified poisoning size. We use the same $N$ and $\alpha$ for all three cases to fairly compare them. Following Jia et al.~\cite{jia2021intrinsic}, we set $N=10^3$ and $\alpha=10^{-3}$ by default. 

$K$ is a  hyperparameter in KNN. We search for the $K$ that achieves the largest $\text{ACPS}$ from the following values: 100, 300, 500, 1,000, 3,000, and 5,000. Following Jia et al.~\cite{jia2022certified}, we use $\ell_1$ distance in KNN. For backdoor attacks, we use a $2\times 2$ white patch as the backdoor trigger. Moreover, following previous work~\cite{gu2017badnets,wang2019neural}, we embed the backdoor trigger at the bottom right corner of an image. 

\subsection{Experimental Results}
\begin{table}[!t]
\caption{Comparing {\LP} and {\FT} for bagging.}
\vspace{-5mm}
\label{bagging-case3}
\begin{center}
\subfloat[Testing accuracy under no attacks]{\begin{tabular}{|c|c|c|}
\hline
           & \LP & \FT \\ \hline
Testing accuracy        & 0.979   & 0.982   \\ \hline
\end{tabular}}

\subfloat[ACPS against attacks]{
\begin{tabular}{|c|c|c|}
\hline
           & \LP & \FT \\ \hline
Data poisoning attacks        & 70.3   & 71.0   \\ \hline
Backdoor attacks        & 69.9   & 70.5   \\ \hline
\end{tabular}}

\subfloat[Computation and storage cost]{
\begin{tabular}{|c|c|c|}\hline
           & \LP & \FT \\ \hline
Training time (seconds) & $1.41\times10^{2}$   &  $1.86\times10^{3}$  \\ \hline
Testing time (seconds)  & $2.26\times 10^{2}$    & $2.97\times 10^{3}$   \\ \hline
Storage (MB)    & $2.15\times 10^{1}$   &  $3.38\times 10^{5}$  \\ \hline
\end{tabular}}
\end{center}
\vspace{-5mm}
\end{table}
\myparatight{Comparing {\LP} with {\FT} for bagging} Table~\ref{bagging-case3} shows the results of bagging in {\LP} and {\FT}  on STL10. First, {\LP} and {\FT} achieve similar testing accuracy and ACPS. In particular, their testing accuracy difference is within 0.003 and their ACPS differences are within 0.6. Second, the computation and storage cost of {\LP} are respectively an order and four orders of magnitude smaller than those of {\FT}. {\LP} takes a smaller training time because it only trains the last fully connected layer for each base classifier; {\LP} takes a smaller testing time because different base classifiers share the same encoder and thus we only need to use it to compute a feature vector for each testing input once; and {\LP} incurs a smaller storage cost because we only need to save the parameters of the last fully connected layer for each base classifier. As {\LP} and {\FT} achieve similar  testing accuracy and ACPS  but {\LP} incurs smaller computation and storage cost, we use {\LP} in {\caseIII} in the following experiments.

\begin{table}[!t]
\caption{Comparing the three cases for bagging.}
\vspace{-5mm}
\label{bagging-results}
\begin{center}
\subfloat[Testing accuracy under no attacks]
{
\begin{tabular}{|c|c|c|c|}
\hline
         & STL10  &  CIFAR10 &  Tiny-ImageNet\\ \hline
\caseI   &   0.352       & 0.358      &      0.071      \\ \hline
\caseII   &   0.559       & 0.566      &   0.072         \\ \hline
\LP &   0.979      &  0.907      &    0.629    \\ \hline
\end{tabular}}

\subfloat[ACPS against data poisoning attacks]
{
\begin{tabular}{|c|c|c|c|}
\hline
         & STL10  &  CIFAR10 &  Tiny-ImageNet\\ \hline
\caseI   &   3.7       & 21.2      &   0.001         \\ \hline
\caseII  &   5.0       & 20.7      &   0.06       \\ \hline
\LP &   70.3      &  359.0      &    19.5    \\ \hline
\end{tabular}}

\subfloat[ACPS against backdoor attacks]{
\label{bagging-backdoor}
\begin{tabular}{|c|c|c|c|}
\hline
         & STL10  & CIFAR10 &  Tiny-ImageNet\\ \hline
\caseI   &  3.7      &  20.9     &  0.001         \\ \hline
\caseII  &   5.0       &  20.3     & 0.06      \\ \hline
\LP &  69.9     &   357.5     &   19.4   \\ \hline
\end{tabular}}
\vspace{-5mm}
\end{center}
\end{table}

\begin{table}[!t]
\caption{Comparing the three cases for KNN.}
\vspace{-5mm}
\label{knn-results}
\begin{center}
\subfloat[Testing accuracy under no attacks]{
\begin{tabular}{|c|c|c|c|}
\hline
         & STL10  & CIFAR10 &  Tiny-ImageNet\\ \hline
\caseI   &   0.194       & 0.229      &   0.029         \\ \hline
\caseII   &   0.230       & 0.327      &   0.038         \\ \hline
\caseIII &   0.963      &  0.865      &    0.583    \\ \hline
\end{tabular}}

\subfloat[ACPS against data poisoning attacks]{
\begin{tabular}{|c|c|c|c|}
\hline
         & STL10  &  CIFAR10 &  Tiny-ImageNet\\ \hline
\caseI   &   36.6       &  187.8     &    10.3        \\ \hline
\caseII  &  14.3        &    164.2   &13.2          \\ \hline
\caseIII &    138.7     &   929.4     &   51.6     \\ \hline
\end{tabular}}

\subfloat[ACPS against backdoor attacks]{
\label{knn-backdoor}
\begin{tabular}{|c|c|c|c|}
\hline
         & STL10  & CIFAR10 &  Tiny-ImageNet\\ \hline
\caseI   &   36.1   &    187.3   & 10.3          \\ \hline
\caseII  &  14.3     &  163.3     &13.2       \\ \hline
\caseIII &    138.5   & 962.2      &  51.3    \\ \hline
\end{tabular}}
\vspace{-5mm}
\end{center}
\end{table}

\myparatight{Comparing the three cases for bagging} Table~\ref{bagging-results} compares the testing accuracy and ACPS of the three cases under the framework of bagging. We have the following observations. First, {\caseIII} achieves significantly higher testing accuracy and ACPS than {\caseI} and {\caseII} among all datasets. This is because base classifiers trained by {\caseIII} are much more accurate than those trained by {\caseI} and {\caseII}, especially when the number of training examples $k$ used to train each base classifier is small. Second, bagging achieves significantly higher ACPS on CIFAR10 than  on the other two datasets. The reason is that CIFAR10 has $9\times$ more training examples than STL10, and thus each subsample created from CIFAR10 is less likely to contain poisoning training examples than STL10 for a given  subsample size $k$ and number of poisoning training examples. The reason why CIFAR10 has higher ACPS  than   Tiny-ImageNet is that the base classifiers on CIFAR10 are more accurate than those on Tiny-ImageNet since Tiny-ImageNet is a more challenging classification task. Third, we find that bagging achieves similar ACPS against  data poisoning attacks and backdoor attacks. This is because a base classifier is very likely to predict the same label for testing inputs with and without the backdoor trigger when its size is small ($2\times2$ in our experiments). 

\myparatight{Comparing the three cases for KNN} Table~\ref{knn-results} shows the testing accuracy and ACPS  of the three cases for KNN. Similar to bagging, we have three observations: 1) {\caseIII} achieves higher testing accuracy and ACPS than {\caseI} and {\caseII}, 2) CIFAR10 has  higher ACPS  than the other two datasets, and 3)   ACPS of each case is similar for data poisoning attacks and backdoor attacks. 

\myparatight{Comparing bagging with KNN in \caseIII} By comparing the results in Table~\ref{bagging-results} and~\ref{knn-results}, we find that bagging achieves higher testing accuracy but lower ACPS than KNN on each dataset. Bagging achieves higher testing accuracy because each base classifier is trained to better fit with the data distribution, while KNN does not have trainable parameters. 
Bagging achieves lower ACPS than KNN because one poisoning training example impacts multiple base classifiers in bagging, but only impacts at most two nearest neighbors of a testing input in KNN. Thus, KNN can tolerate more poisoning training examples than bagging. 
We also note that, compared with KNN, bagging is a more general framework as we can use an arbitrary learning algorithm to train each base classifier.

\begin{figure}[!t]
    \centering
    \subfloat[Testing accuracy]{\includegraphics[width=0.4\textwidth]{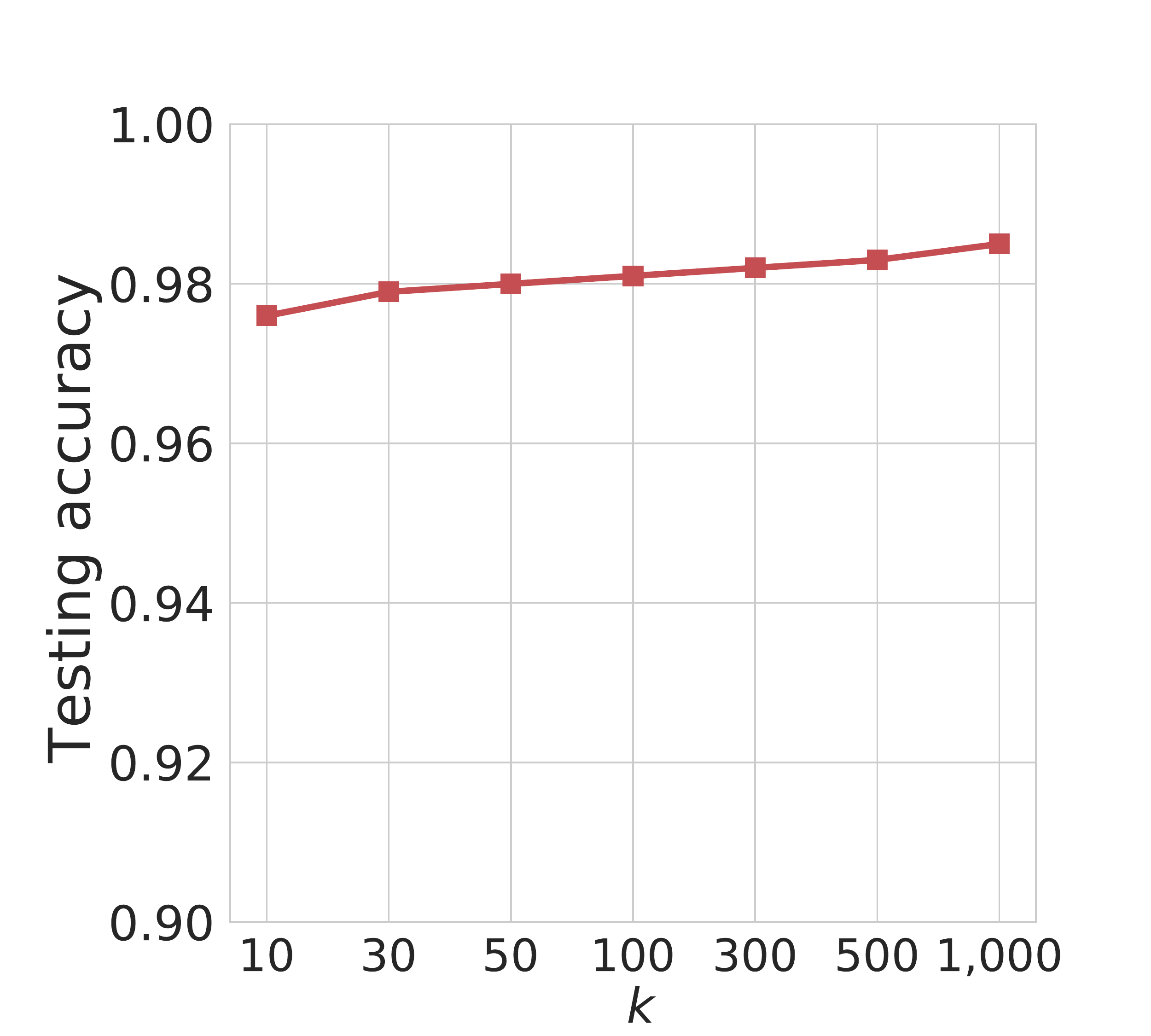}}
    \subfloat[ACPS]{ 
    \includegraphics[width=0.4\textwidth]{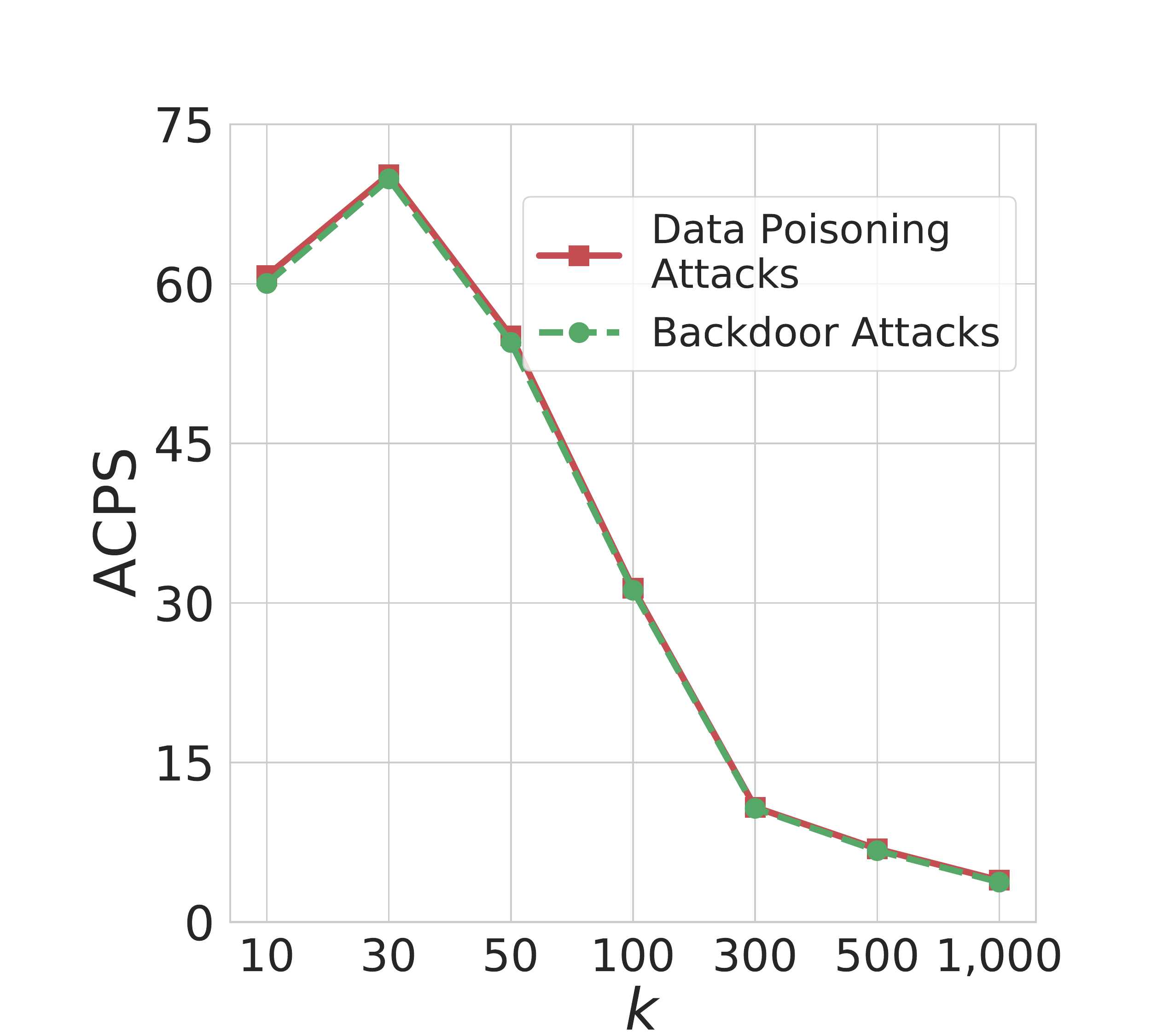}}
    \caption{Impact of $k$ on the testing accuracy of bagging under no attacks and ACPS of bagging against data poisoning and backdoor attacks.}
    \label{poison-figure-impact-k}
\end{figure}

\myparatight{Impact of $k$, $N$, and $\alpha$ on bagging in \caseIII} Figure~\ref{poison-figure-impact-k} shows the impact of $k$ on testing accuracy and ACPS of bagging in {\caseIII} on STL10 dataset. We find that testing accuracy increases as $k$ increases. This is because the base classifiers are more accurate when $k$ is larger as  more training examples are used to train each base classifier. Moreover, we find that  ACPS first increases and then decreases as $k$ increases. The reason is that base classifiers are less accurate when $k$ is small and are more likely to be trained on poisoning training examples when $k$ is large. Figure~\ref{poison-figure-impact-N} in Appendix shows the impact of $N$ on the testing accuracy and ACPS of bagging. We find that $N$ has a negligible impact on testing accuracy. The reason is that a small $N$ can already well estimate the predicted label of bagging for a testing input. Moreover, ACPS increases as $N$ increases. The reason is that a larger $N$ results in tighter estimation of the lower and upper bounds of label probabilities, which leads to a larger certified poisoning size.  Figure~\ref{poison-figure-impact-alpha} in Appendix shows the impact of $\alpha$ on ACPS of bagging in {\caseIII}. Note that testing accuracy of bagging under no attacks does not depend on $\alpha$. We observe that ACPS increases as  $\alpha$ increases. This is because a larger $\alpha$ results in tighter estimation of label-probability bounds. 

\begin{figure}[!t]
    \centering
        \subfloat[Testing accuracy]{\includegraphics[width=0.4\textwidth]{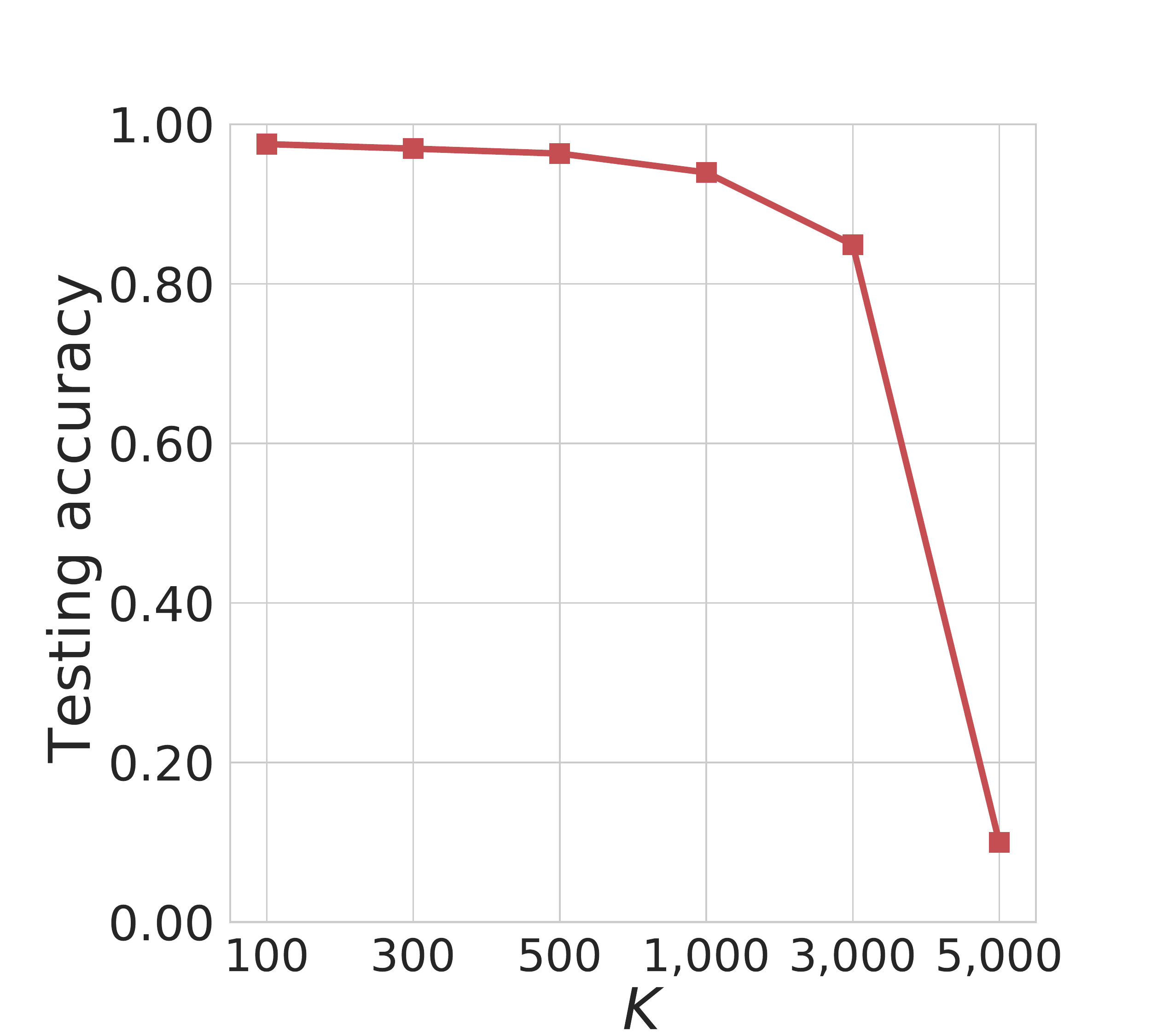}\label{impact-of-k-on-testing-accuracy-of-knn}}
    \subfloat[ACPS]{\includegraphics[width=0.4\textwidth]{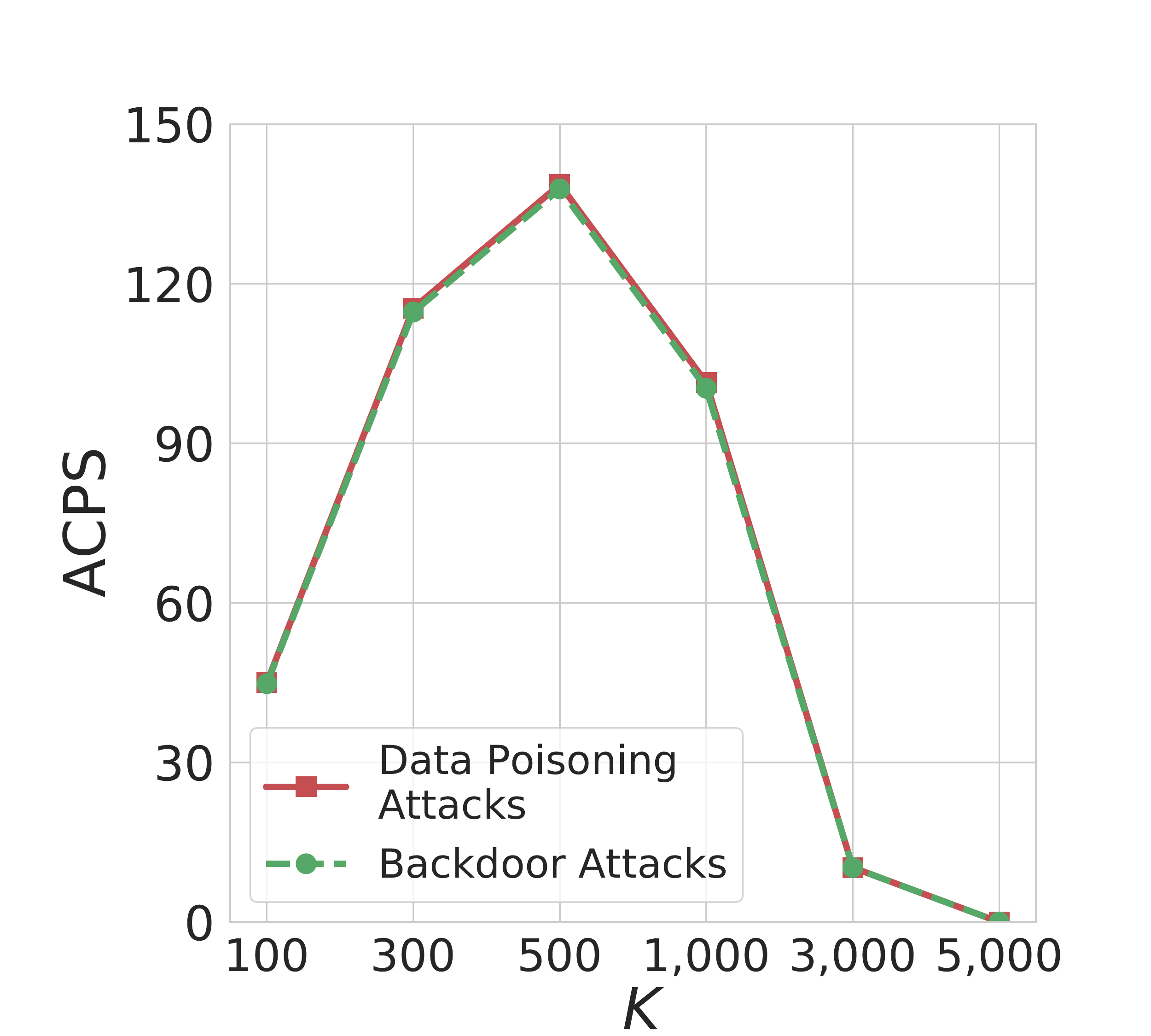}}
    \caption{Impact of $K$ on the (a) testing accuracy of KNN under no attacks, and (b) ACPS of KNN against data poisoning and backdoor attacks.}
    \label{poison-figure-impact-k-knn}
\end{figure}

\myparatight{Impact of   $K$ on KNN in \caseIII} Figure~\ref{poison-figure-impact-k-knn} shows the impact of $K$ on the testing accuracy under no attacks and ACPS of KNN against data poisoning and backdoor attacks. First, we find that testing accuracy decreases as $K$ increases. This is because the decision boundary of KNN classifier is smoother when $K$ is larger. Second, we observe that ACPS of KNN first increases and then decreases as $K$ increases. The reasons are as follows. When $K$ is too small, the gap between $K^a_{\mathbf{x}}$ and $K^b_{\mathbf{x}}$ is small and thus the certified poisoning size is small based on Equation~\ref{formal_cps_knn}. When $K$ is too large, KNN is less accurate compared with a smaller $K$ as shown in Figure~\ref{impact-of-k-on-testing-accuracy-of-knn}.

\myparatight{Summary} We observe that a pre-trained encoder makes bagging and KNN more accurate under no attacks and more secure against data poisoning and backdoor attacks. Compared with bagging, KNN is less accurate under no attacks but achieves a stronger certified security guarantee.  Bagging with linear probing and bagging with fine-tuning achieve similar testing accuracy under no attacks and certified security guarantees against attacks, but bagging with linear probing is orders of magnitude more space and time efficient. 
\section{Security against Adversarial Examples}
\label{sec-adversarial}
\subsection{Related Work}
\myparatight{Adversarial examples} An attacker can add carefully crafted noise to a testing input such that a classifier predicts an arbitrary or attacker-chosen incorrect label for it~\cite{szegedy2013intriguing,carlini2017towards}.  Such testing input with carefully crafted noise is called \emph{adversarial example}. Adversarial examples pose severe security threats to classifiers, especially for security-critical applications such as autonomous driving. 

\myparatight{Certified defenses against adversarial examples}
To defend against adversarial examples, both empirical defenses~\cite{papernot2016distillation,xu2018feature,goodfellow2014explaining,madry2018towards,cao2017mitigating} and certified defenses~\cite{cohen2019certified,huang2017safety,pinot2019theoretical,salman2019provably,jia2019certified} were proposed. Empirical defenses do not have formal security guarantees against adversarial examples, and thus they are often broken by adaptive, advanced attacks~\cite{carlini2017adversarial,athalye2018obfuscated}. Therefore, we focus on certified defenses in this work. A certified defense can prove a classifier's predicted label for a testing input does not change when the magnitude (e.g., measured by $\ell_2$-norm) of the noise added to the testing input is smaller than a threshold (called \emph{certified radius}). Formally, given a testing input $\mathbf{x}$ and a classifier $g$, a certified defense can provide the following guarantee:
\begin{align}
    g(\mathbf{x}) = g(\mathbf{x}+\delta), \forall \delta \text{ s.t. } \|\delta\|_{2}<R_{\mathbf{x}},
\end{align}
where $R_{\mathbf{x}}$ is the certified radius for the testing input $\mathbf{x}$.
Among existing certified defenses, randomized smoothing~\cite{cohen2019certified} is state-of-the-art  since it is applicable to arbitrary classifiers and scalable to large neural networks. Next, we  discuss it in more details.

\myparatight{Randomized smoothing~\cite{cohen2019certified}} 
Given a testing input $\mathbf{x}$, randomized smoothing first creates a randomized input via adding isotropic Gaussian noise to it. Then, randomized smoothing uses an arbitrary classifier $f$ (called \emph{base classifier}) to predict a label for the randomized input. Due to randomness in the randomized input, the predicted label is also random. To capture such randomness, we define label probability $p_i$ as the probability that the base classifier $f$ predicts label $i$ for the randomized input, where $i=1,2,\cdots,c$. Formally, we define $p_i = \Pr(f(\mathbf{x} + \mathcal{N}(0, \sigma^{2} I)) = i)$, where $\mathcal{N}(0, \sigma^{2} I)$ is isotropic Gaussian noise with standard deviation $\sigma$. Given the label probabilities, randomized smoothing builds a \emph{smoothed classifier}  $g$ which predicts the label with the largest label probability for  $\mathbf{x}$. Formally, we have: $g(\mathbf{x}) = \argmax_{i=1,2,\cdots,c} p_i$. 

Cohen et al.~\cite{cohen2019certified} proposed a Monte-Carlo method to compute $g(\mathbf{x})$ and the certified radius $R_{\mathbf{x}}$. Roughly speaking, they first generate $M$ noise from  $\mathcal{N}\left(0, \sigma^{2} I\right)$, then add each of them to $\mathbf{x}$ to create $M$ noisy inputs, and finally use the base classifier $f$ to predict a label for each noisy input. Given the predicted labels, they count the number of noisy inputs that are predicted as label $i$ ($i=1,2,\cdots,c$). The label with the largest count (denoted as $a$) is predicted as $g(\mathbf{x})$. Given the count of the label $a$, Cohen et al. proposed to use one-sided Clopper-Pearson method~\cite{clopper1934use} to estimate a lower bound of the label probability $p_a$, which we denote as $\underline{p_a}$. Cohen et al. showed that the certified radius $R_{\mathbf{x}}$ can be computed as follows:
\begin{align}
\label{randomized_smoothing_certified_radius}
R_{\mathbf{x}}=\frac{\sigma}{2}(\Phi^{-1}(\underline{p_{a}})-\Phi^{-1}(1-\underline{p_{a}})),
\end{align}
where $\Phi^{-1}$ is the inverse of the standard Gaussian cumulative distribution function. Due to probabilistic errors in estimating the label-probability lower bound $\underline{p_a}$, the outputted certified radius  $R_{\mathbf{x}}$ may be incorrect with a probability $\alpha$ (called \emph{confidence level}), which can be set to be arbitrarily small.  

\myparatight{Limitations of randomized smoothing} The key limitations of randomized smoothing are that: 1) it sacrifices testing accuracy  under no attacks, and 2) its certified radius is small. We aim to study whether and when a pre-trained encoder can mitigate these limitations of randomized smoothing. 

\subsection{Experimental Setup}
\myparatight{Training base classifiers}
Given a training dataset, we train a base classifier in {\caseI}, {\caseII}, {\LP}, or {\FT}.  Cohen et al.~\cite{cohen2019certified} showed that the smoothed classifier achieves better accuracy under no attacks and certified radius if the isotropic Gaussian noise is added to the training inputs when training the base classifier. Therefore, we add random isotropic Gaussian noise to the training inputs in each epoch when training a base classifier in each case. 

\myparatight{Evaluation metrics} We use \emph{testing accuracy under no attacks (testing accuracy)} and \emph{average certified radius (ACR)} as the evaluation metrics. In particular, testing accuracy measures the fraction of testing examples  that are correctly classified by a smoothed classifier. ACR measures the average certified radius of the testing examples that are correctly classified by a smoothed classifier. Formally, given a testing dataset $\mathcal{D}_{te}$, ACR can be computed as follows:
\begin{align}
ACR = \frac{\sum_{(\mathbf{x}_{i}, y_i) \in \mathcal{D}_{te} } \mathbb{I}(a_{\mathbf{x}_i}=y_i) \cdot R_{\mathbf{x}_i}}{\sum_{(\mathbf{x}_{i}, y_i) \in \mathcal{D}_{te} } \mathbb{I}(a_{\mathbf{x}_i}=y_i)} ,
\end{align}
where $a_{\mathbf{x}_i}$ and $R_{\mathbf{x}_i}$ are respectively the predicted label, certified radius for the testing input $\mathbf{x}_i$, and $\mathbb{I}$ is an indicator function.

\myparatight{Hyperparameter search and parameter setting} 
$\sigma$ achieves a tradeoff between testing accuracy and ACR of a smoothed classifier as shown in Figure~\ref{adv-figure-impact-sigma-accuracy-ACR}. In particular,  testing accuracy decreases and ACR increases as $\sigma$ increases. As a result, $\sigma$ would be $\infty$ (or $0$) if we search for it to achieve the largest ACR (or testing accuracy) in each case, which leads to $\infty$ (or $0$) ACR based on Equation~\ref{randomized_smoothing_certified_radius} and makes   comparison of ACR across the three cases meaningless. In response, we compare the three cases by following previous work on how to compare different smoothed classifiers~\cite{cohen2019certified,jia2019certified}. In particular, we set $\sigma$ to be a small value (i.e., $0.25$) for {\caseI} and {\caseII} and then search for $\sigma$ in {\caseIII} such that the testing accuracy of the smoothed classifier in {\caseIII} is similar to or higher than those in {\caseI} and {\caseII}. In particular, we search among the following $\sigma$ in {\caseIII}: 0.25, 0.50, 0.75, and 1.00. Note that for each case, we still search for the combination of  number of training epochs and learning rate that achieves the largest ACR when training a base classifier, following the description in Section~\ref{sec:experimentalsetup}. 
In the testing phase, $M$ and $\alpha$ are two parameters used to estimate the predicted label and label-probability bound for a testing input. We set them to be the same in the three cases for fair comparison. Following Cohen et al.~\cite{cohen2019certified}, we set $M=10^4$ and $\alpha=10^{-3}$ by default.

\begin{table}[!t]
\caption{Comparing {\LP} and {\FT} in randomized smoothing.}
\vspace{-5mm}
\label{table_rs_our_new_methods}
\begin{center}
\begin{tabular}{|c|c|c|}
\hline
& {\LP}        & {\FT} \\ \hline
 Testing accuracy    &    0.390   &  0.612     \\ \hline
  ACR    &     0.815   &  1.482     \\ \hline
\end{tabular}
\end{center}
\vspace{-5mm}
\end{table}

\begin{table}[!t]
\vspace{4mm}
\caption{Comparing three cases for randomized smoothing.}
\label{rs-comparison-results}
\begin{center}
\vspace{-6mm}
\subfloat[Testing accuracy under no adversarial examples]{\begin{tabular}{|c|c|c|c|}
\hline
         & STL10 &  CIFAR10  & Tiny-ImageNet\\ \hline
\caseI   &  0.272    &  0.312   &  0.036     \\ \hline
\caseII  &  0.610   &  0.754   &  0.390      \\ \hline
\FT &  0.612    &  0.806      &  0.418  \\ \hline
\end{tabular}}

\vspace{-2mm}

\subfloat[ACR against adversarial examples]{\begin{tabular}{|c|c|c|c|}
\hline
         & STL10 &  CIFAR10  & Tiny-ImageNet\\ \hline
\caseI   &  0.295    &  0.411   &  0.315      \\ \hline
\caseII  &  0.625    &  0.518   &  0.482      \\ \hline
\FT &  1.482    &  0.593    & 0.767      \\ \hline
\end{tabular}}
\end{center}
\vspace{-9mm}
\end{table}

\vspace{-2mm}
\subsection{Experimental Results}
\vspace{-3mm}
\myparatight{Comparing {\LP} with {\FT}} Table~\ref{table_rs_our_new_methods} compares testing accuracy and ACR of  {\LP} and {\FT} on STL10 dataset. Our results show that {\FT} achieves better testing accuracy and ACR than {\LP}. The reason is that {\FT} updates both the encoder and the last fully connected layer using training inputs added with Gaussian noise when training the base classifier. Thus, the base classifier trained by {\FT} is more likely to correctly  classify the testing inputs added with Gaussian noise. Therefore, unless otherwise mentioned, we use {\FT} in {\caseIII} in the following experiments.

\myparatight{Comparing the three cases} Table~\ref{rs-comparison-results} compares the testing accuracy and ACR of the three cases. We have the following observations. First, {\FT} achieves the highest ACR while its testing accuracy is similar to or higher than those of {\caseI} and {\caseII}. In other words, {\FT} can offer a stronger certified security guarantee against adversarial examples than {\caseI} and {\caseII} without sacrificing  testing accuracy under no attacks. The reason is that a pre-trained encoder can produce high-quality features and thus the base classifier in {\FT} is more likely to correctly classify the testing inputs  with Gaussian noise. Second, we find that {\FT} achieves significantly higher ACR on STL10 than the other two datasets. We suspect the reason is that the number of training examples in STL10 is much smaller than the other two datasets and thus the performance improvement brought by high-quality features produced by the pre-trained encoder is more significant. Another observation is that {\caseII} achieves higher testing accuracy and ACR than {\caseI}. The reason is that deep neural network is more powerful in generalization than a linear classifier.

\begin{figure}[!t]
\vspace{-5mm}
    \centering
    \subfloat[Testing accuracy]{ 
    \includegraphics[width=0.4\textwidth]{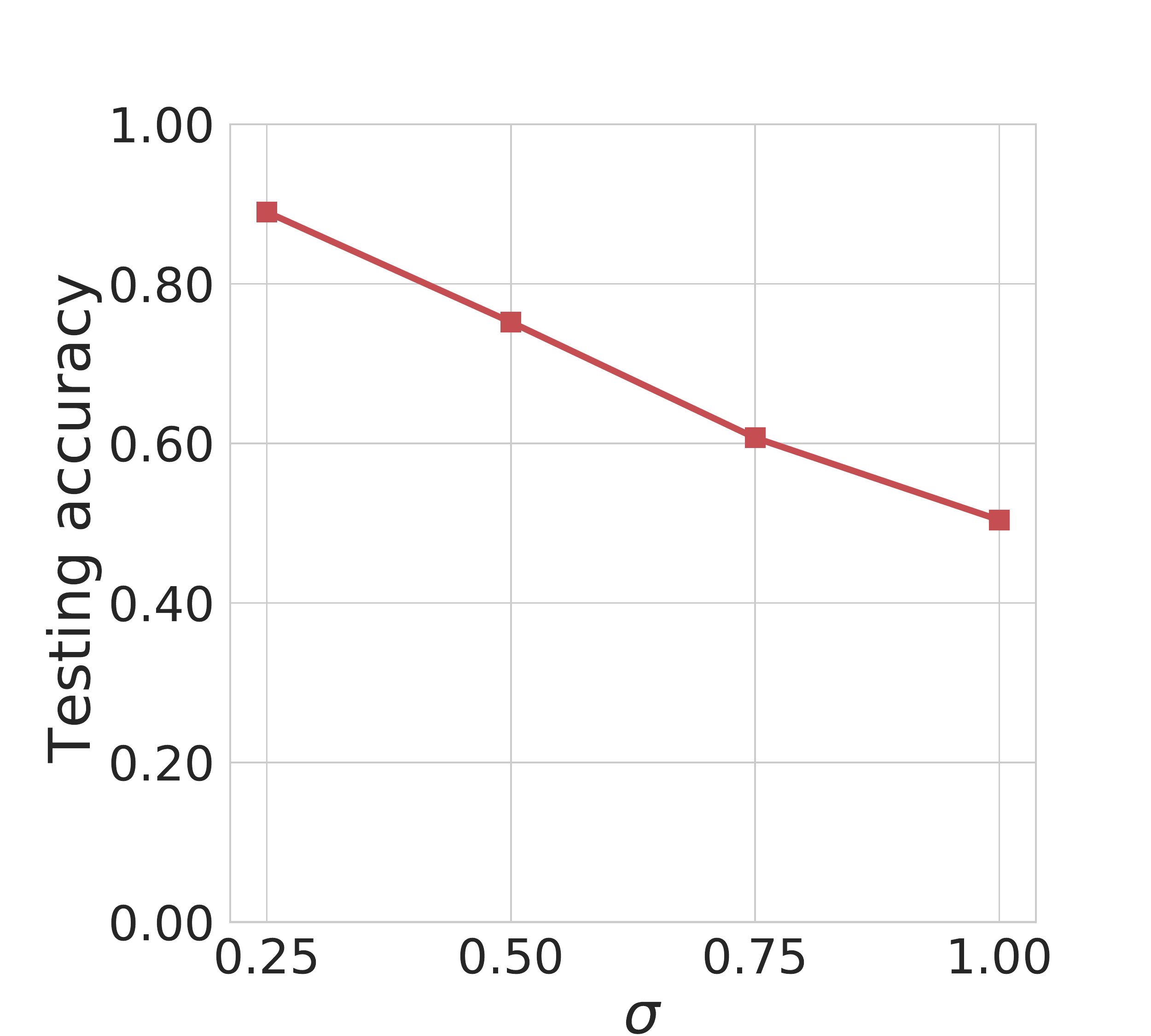}\label{adv-figure-impact-sigma-accuracy}}
    \subfloat[ACR]{ \includegraphics[width=0.4\textwidth]{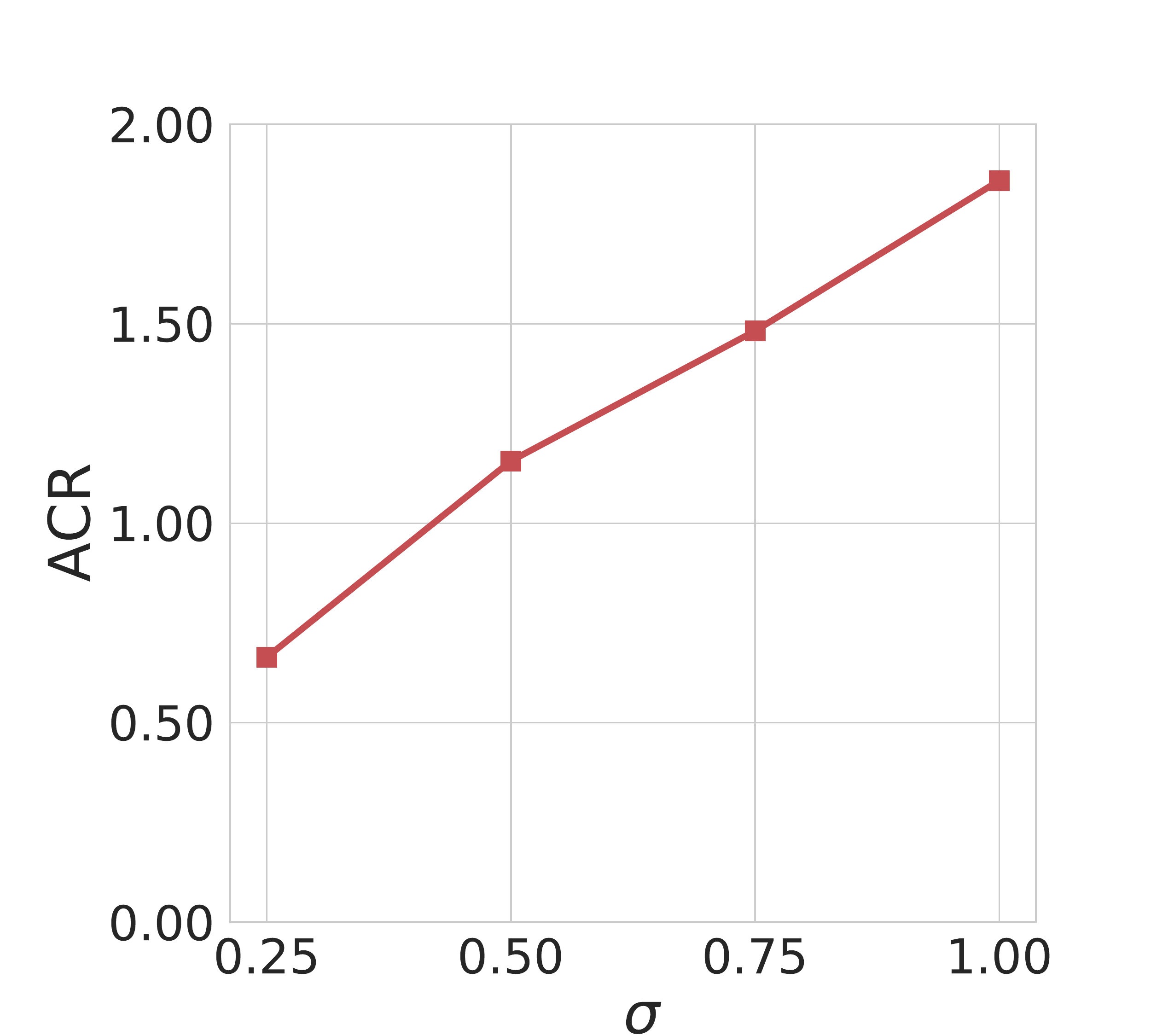}\label{adv-figure-impact-sigma-ACR} }
    \caption{Impact of $\sigma$ on the testing accuracy and ACR of randomized smoothing.}
    \label{adv-figure-impact-sigma-accuracy-ACR}
\end{figure}

\myparatight{Impact of $\sigma$, $M$, and $\alpha$ in {\FT}} Figure~\ref{adv-figure-impact-sigma-accuracy-ACR} shows the impact of $\sigma$ on  testing accuracy and ACR of the smoothed classifier in {\FT} on STL10. First, we find that  testing accuracy decreases as $\sigma$ increases, which is because adding a larger noise to testing inputs makes it harder to correctly classify them. Second, we find that ACR increases as $\sigma$ increases. The reason is that adding a larger noise to a testing input requires a larger adversarial perturbation to manipulate its predicted label. Figure~\ref{adv-figure-N-accuracy-ACR} in Appendix shows the impact of $M$ on  testing accuracy and ACR of the smoothed classifier. We find that $M$ has a negligible impact on testing accuracy. This is because a small $M$ can already provide  good estimations of the labels predicted by the smoothed classifier for the testing inputs. We also observe that ACR increases as $M$  increases. The reason is that a larger $M$ produces a tighter  label-probability lower bound $\underline{p_a}$ and thus a larger certified radius. Figure~\ref{adv-figure-impact-alpha-ACR} in Appendix shows the impact of $\alpha$ on ACR of the smoothed classifier. We find that ACR increases as $\sigma$ increases. The reason is that a larger $\alpha$ leads to a tighter $\underline{p_a}$. We note that $\alpha$ does not impact testing accuracy.

\myparatight{Training with vs. without Gaussian noise}  Table~\ref{rs_caseIII_with_vs_without_noise} compares the testing accuracy and ACR of {\FT} when training the base classifiers with or without adding Gaussian noise to the training inputs. We find training with noise achieves significantly higher testing accuracy and ACR than training without noise. The reason is that,  in training with Gaussian noise, both training and testing data of the base classifier have the same distribution and thus the base classifier is more accurate at classifying testing inputs with Gaussian noise.

\begin{table}[!t]
\caption{Training with vs. without Gaussian noise.}
\vspace{-5mm}
\label{rs_caseIII_with_vs_without_noise}
\begin{center}
\begin{tabular}{|c|c|c|c|}
\hline
 & Training w/o noise & Training with noise \\ \hline
Testing accuracy & 0.358 & 0.612 \\ \hline
ACR & 0.729 & 1.482 \\ \hline
\end{tabular}
\vspace{-5mm}
\end{center}
\end{table}

\myparatight{Summary} We observe that a pre-trained encoder can improve both the testing accuracy under no attacks and certified security guarantees against adversarial examples of randomized smoothing. The smoothed classifier  can achieve better testing accuracy and certified security guarantees  when training a base classifier via fine-tuning and adding  Gaussian noise to the training inputs. 

\section{Training with Differential Privacy}
\label{sec-dp}
\subsection{Related Work} 
\myparatight{Inference attacks to classifiers}  Classifiers are vulnerable to various inference attacks such as model inversion attacks~\cite{fredrikson2015model}, membership inference attacks~\cite{shokri2017membership},  and property inference attacks~\cite{ateniese2015hacking,ganju2018property}. In particular, given either a black-box or white-box access to a classifier, an attacker can infer various information about its training data. These attacks cause severe privacy concerns  when the training data is privacy-sensitive, e.g., medical images. 

\myparatight{Differentially private classifiers} 
To defend against inference attacks, many studies~\cite{abadi2016deep,song2013stochastic,bassily2014private,chaudhuri2011differentially,kifer2012private,liu2020intrinsic} extend differential privacy~\cite{dwork2014algorithmic}, a rigorous privacy notation, to classifier training. Roughly speaking, a learning algorithm is differentially private if the distribution of the model parameters of  its learnt classifier does not change much when   one training example is added or deleted in the training dataset. Suppose we have a randomized learning algorithm $\mathcal{A}$ whose input is a training dataset $\mathcal{D}$ and whose output is a classifier $f\in \mathcal{F}$, where $\mathcal{F}$ is the space of classifier. We denote by $\mathcal{A}(\mathcal{D})$  the classifier learnt by $\mathcal{A}$ on $\mathcal{D}$.  Two training datasets $\mathcal{D}$ and $\mathcal{D}'$ are said to be adjacent if and only if $\mathcal{D}$ and $\mathcal{D}'$ differ by exactly one training example. Formally, we have the following definition for $(\epsilon,\delta)$-differential privacy:
\begin{definition}[$(\epsilon,\delta)$-differential privacy~\cite{dwork2014algorithmic}] A randomized learning algorithm $\mathcal{A}$ is $(\epsilon,\delta)$-differentially private if for any $\mathcal{R} \subseteq \mathcal{F}$ and for any two adjacent training datasets $\mathcal{D}$ and $\mathcal{D}^{\prime}$, $\mathcal{A}$ satisfies the following: $\Pr[\mathcal{A}(\mathcal{D}) \in \mathcal{R}] \leq e^{\epsilon} \Pr[\mathcal{A}(\mathcal{D}^{\prime}) \in \mathcal{R}] + \delta$.
\end{definition}

A smaller $\epsilon$ indicates a stronger privacy guarantee. $\epsilon=\infty$ means no privacy guarantee, which is equivalent to training a non-private classifier. Many differentially private  learning algorithms~\cite{abadi2016deep,song2013stochastic,bassily2014private,chaudhuri2011differentially,kifer2012private,liu2020intrinsic} have been proposed. Their key idea is to introduce  randomness to the training process. For instance, they  may add random noise to the training examples~\cite{jordon2018pate}, gradients~\cite{abadi2016deep,song2013stochastic,bassily2014private} used to update the parameters of a classifier, and/or loss functions~\cite{chaudhuri2011differentially,kifer2012private}. Among them, DP-SGD~\cite{abadi2016deep} is state-of-the-art and is applicable to deep neural networks. Roughly speaking, DP-SGD clips the gradients and adds Gaussian noise to them before using them to update the parameters of a classifier during training. 
 
\myparatight{Limitations of differentially private classifiers} A key limitation of  differentially private learning algorithms is that they substantially sacrifice the testing accuracy of the classifiers. In particular, a differentially private classifier  is (much) less accurate than its non-private counterpart. Such limitation is a key obstacle for    deploying differentially private classifiers in the real world. We aim to measure whether a pre-trained encoder can mitigate such limitation. 

\vspace{-2mm}
\subsection{Experimental Setup}
\vspace{-2mm}
\myparatight{Training differentially private classifiers} We use DP-SGD~\cite{abadi2016deep} to train classifiers that achieve $(\epsilon,\delta)$-differential privacy. We adopt the public implementation~\cite{yousefpour2021opacus} of DP-SGD in our experiments. For {\caseI} and {\caseII}, we train classifiers with DP-SGD from scratch. For {\LP}, we use DP-SGD to train the parameters of the last fully connected layer of the classifier, while for {\FT}, we  use DP-SGD to update the pre-trained encoder and the last fully connected layer. 

\myparatight{Evaluation metric}
We use testing accuracy to evaluate the performance of a differentially private classifier. In particular, given a testing dataset and a classifier, testing accuracy is the fraction of the testing examples that are correctly classified. 

\myparatight{Hyperparameter search and parameter setting} 
Following previous works~\cite{dwork2014algorithmic,abadi2016deep}, we set $\delta=10^{-5}$. Moreover,  unless otherwise mentioned, we set $\epsilon=1.0$ to achieve a strong privacy guarantee. Given $\epsilon$ and $\delta$, we use grid search to find the best combination of the number of training epochs and learning rate for DP-SGD in each case, following the hyperparameter search process  in Section~\ref{sec:experimentalsetup}. Specifically, we find the best hyperparameter setting of DP-SGD that  
 achieves the largest accuracy on the validation examples. 

\vspace{-2mm}
\subsection{Experimental Results}
\vspace{-2mm}
\begin{figure}[!t]
    \centering
    \vspace{-5mm}
    \subfloat[]{\includegraphics[width=0.4\textwidth]{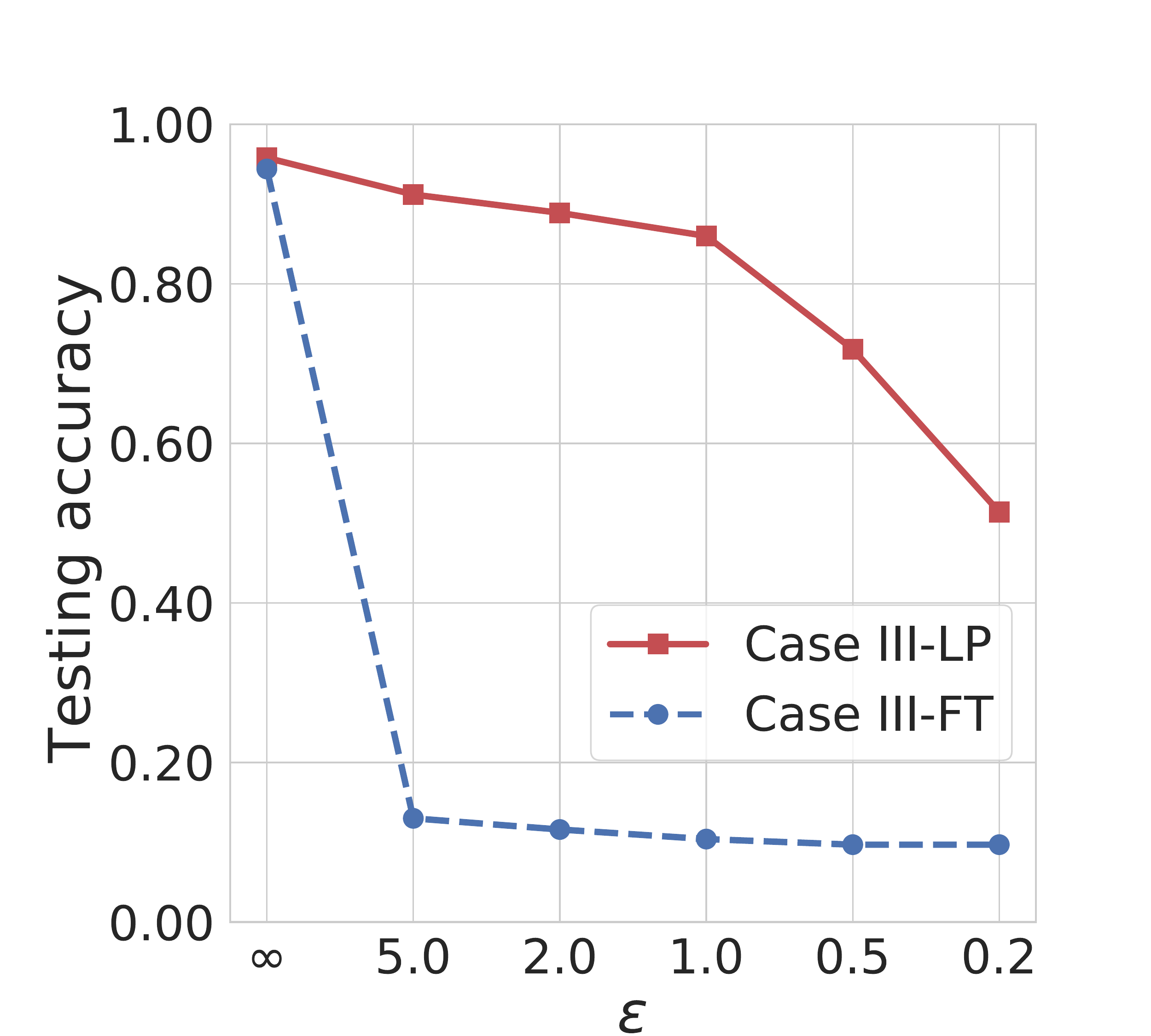}\label{dp-figure-impact-lp-ft}}
    \subfloat[]{\includegraphics[width=0.4\textwidth]{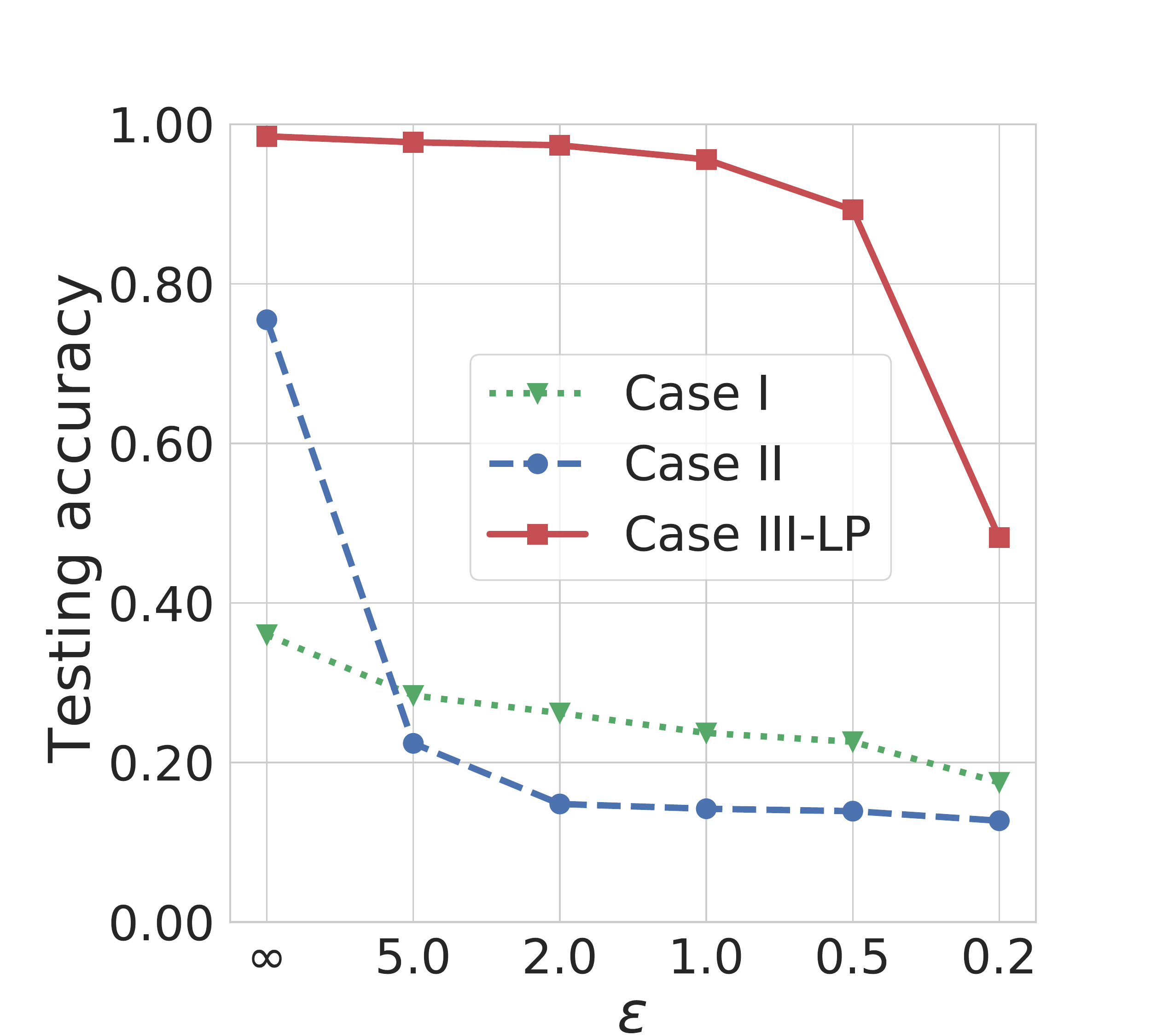}\label{dp-figure-impact-epsilon-testing-acc}}
    \caption{(a) Comparing  testing accuracy of {\LP} and {\FT} as $\epsilon$ varies. (b) Comparing  testing accuracy of the three cases as $\epsilon$ varies.} 
    \label{dp-figure-impact-epsilon}
\end{figure}

\vspace{-1mm}
\myparatight{Comparing {\LP} with {\FT}}
Figure~\ref{dp-figure-impact-lp-ft} compares the testing accuracy of {\LP} and {\FT} on STL10 dataset when $\epsilon$ decreases from $\infty$ (no privacy guarantee) to 0.2 (strong privacy guarantee). In these experiments, we use the image encoder~\cite{chen2020simple} pre-trained by Google on the ImageNet dataset, which is publicly available  on GitHub~\cite{google_github}.
We did not use the OpenAI's CLIP encoder because its neural network architecture is a vision transformer, whose attention pooling layer is not supported by the public implementation of DP-SGD. In particular, {\FT} uses DP-SGD to update both the pre-trained encoder and the last fully connected layer. Since the public implementation of DP-SGD does not support vision transformer, we were not able to perform experiments on DP-SGD for {\FT} using the CLIP encoder. We note that we use the Google's pre-trained encoder only in the experiments in Figure~\ref{dp-figure-impact-lp-ft}, and we perform all other experiments in {\caseIII}  using CLIP.

We observe that the testing accuracy of {\FT} drops sharply while the testing accuracy of {\LP} drops slowly as $\epsilon$ decreases. In other words, {\LP} achieves much higher testing accuracy than {\FT} under differential privacy. This is because {\FT} adds larger noise to the gradients as it updates both the pre-trained encoder and the last fully connected layer during training.  
Our results indicate that  {\LP} should be used to train a differentially private classifier when a pre-trained encoder is available. Therefore, in the following, we focus on LP in {\caseIII}. 

\vspace{-1mm}
\myparatight{Comparing the three cases} Table~\ref{dp-results-compare-all} shows the testing accuracy of the classifiers in the three cases under strong privacy guarantee on STL10 dataset. We have two observations from the results. First, we observe that {\caseIII} achieves significantly higher testing accuracy than the other two cases. For instance, the testing accuracies are respectively 0.237, 0.142, and 0.956 for {\caseI}, {\caseII}, and {\caseIII} on STL10 dataset. The reason is that the pre-trained encoder in {\caseIII} can extract high-quality features, enabling DP-SGD to train more accurate, diffrentially private classifiers. 
We note that our observation is consistent with prior work~\cite{tramer2020differentially}, which also found that a pre-trained encoder can improve the accuracy of a differentially private classifier. But our evaluation is more comprehensive since we compare training differentially private classifiers using linear probing and fine-tuning when having a pre-trained encoder. Second, {\caseII} achieves much higher testing accuracy than {\caseI} under no privacy guarantee (please see Table~\ref{comparison-non-adversarial}), but we find that the testing accuracy of {\caseI} is  similar to or  even higher than that of {\caseII} under strong privacy guarantee. The reason is that a deep neural network classifier in {\caseII} has much more  parameters and thus DP-SGD introduces larger  noise during training, which overwhelms the expressiveness advantages of a deep neural network. Our observation indicates that, without a pre-trained encoder, we can train a  simple linear classifier instead of a complex deep neural network to (possibly) achieve better accuracy under strong privacy guarantee. 

\begin{table}[!t]
\caption{Comparing the testing accuracy of the three cases under $(1.0,10^{-5})$-differential privacy.}
\vspace{-5mm}
\label{dp-results-compare-all}
\begin{center}
\begin{tabular}{|c|c|c|c|}
\hline
         & STL10 &  CIFAR10  & Tiny-ImageNet\\ \hline
{\caseI}   &   0.237      &  0.344    &    0.020    \\ \hline
{\caseII}  &   0.142      &  0.338     &   0.021    \\ \hline
{\LP} &  0.956       &  0.918     &   0.613    \\ \hline
\end{tabular}
\end{center}
\vspace{-5mm}
\end{table}

\vspace{-1mm}
\myparatight{Impact of $\epsilon$}
Figure~\ref{dp-figure-impact-epsilon-testing-acc} shows the impact of $\epsilon$ on the testing accuracy for the three cases on STL10 dataset. First, we observe that the testing accuracy of each case decreases as $\epsilon$ decreases (i.e., privacy guarantee becomes stronger). 
The reason is that a smaller $\epsilon$  requires adding larger noise  during  training, leading to less accurate classifiers. Second, the testing accuracy of {\LP} decreases slowly when  $\epsilon > 1$, after which it decreases quickly. For instance, compared to {\LP} with $\epsilon=\infty$ (i.e., no privacy guarantee), the testing accuracy of {\LP} with $\epsilon=1$ (strong privacy guarantee) decreases only by 0.029.  However, the testing accuracy of {\caseII} decreases sharply (i.e., by 0.703) when $\epsilon$ decreases from $\infty$ to 5. 
Third, {\caseI} achieves larger testing accuracy than {\caseII} when $\epsilon$ is small (e.g., $\epsilon\leq 5$), indicating that a simple linear classifier is preferred over a complex deep neural network  when   a pre-trained encoder is not available but strong privacy guarantee is desired.

\vspace{-1mm}
\myparatight{Summary} We observe that a pre-trained encoder substantially improves the accuracy of a differentially private classifier. 
When a pre-trained encoder is not available,  a  differentially private linear classifier can be more accurate than a  differentially private  deep neural network when strong privacy guarantee is desired. 
\vspace{-2mm}

\section{Exact Machine Unlearning}
\label{sec-unlearn}
\vspace{-2mm}
\subsection{Related Work}
\vspace{-2mm}
\myparatight{The right to be forgotten} The right to be forgotten refers to the right to have a user's private data removed from a service. Various regulations and laws, e.g., the European Union's General Data Protection Regulation (GDPR)~\cite{voigt2017eu}, have established such right for users. In the context of machine learning, a user may request deleting his/her data from the training dataset and removing its effect on a classifier after the classifier has been trained.   

\myparatight{Exact unlearning} \emph{Machine unlearning}~\cite{cao2015towards} aims to achieve the right to be forgotten in the context of machine learning. Specifically, given a set of training examples to be forgotten/unlearnt (denoted as $\mathcal{D}_{unlearn}$),  machine unlearning aims to remove the effect of these training examples on a  classifier. Based on the completeness of unlearning, machine unlearning can be categorized into \emph{exact unlearning}~\cite{cao2015towards,ginart2019making,bourtoule2021machine} and \emph{approximate unlearning}~\cite{wu2020deltagrad,guo2019certified}. Roughly speaking, exact unlearning completely removes the effect of the training examples to be unlearnt on the classifier, while approximate unlearning does not. Therefore, we focus on exact unlearning in this work because it provides strong privacy guarantees for the right to be forgotten.  Following Cao \& Yang~\cite{cao2015towards}, we formally define exact unlearning as follows:

\begin{definition}
[Exact Unlearning] Suppose we have a training dataset $\mathcal{D}$, a set of training examples $\mathcal{D}_{unlearn}$ to be unlearnt, a randomized learning algorithm $\mathcal{A}$, which takes a training dataset as input and outputs a classifier. 
We denote an unlearning algorithm as $\mathcal{U}$, which takes $\mathcal{A}$, $\mathcal{D}$, $\mathcal{A}(\mathcal{D})$, and  $\mathcal{D}_{unlearn}$ as input and outputs a classifier. $\mathcal{A}(\mathcal{D}-\mathcal{D}_{unlearn})$ is the classifier learnt by $\mathcal{A}$ on the remaining training examples $\mathcal{D}-\mathcal{D}_{unlearn}$. Since $\mathcal{A}$ is randomized, the learnt classifier is also randomized in the space of classifiers. We denote by  $\mathbb{F}$ the probability distribution of the classifier $\mathcal{A}(\mathcal{D}-\mathcal{D}_{unlearn})$ in the classifier space. Moreover, we denote by $\mathbb{F}^{\prime}$ the probability distribution of the classifier outputted by the unlearning algorithm $\mathcal{U}$. We call $\mathcal{U}$ exact unlearning for $\mathcal{A}$ if and only if $\mathbb{F} = \mathbb{F}^{\prime}$ holds for arbitrary $\mathcal{D}$ and $\mathcal{D}_{unlearn}$.
\end{definition}

The most popular exact unlearning method is to retrain a classifier from scratch on the remaining training examples, i.e., $\mathcal{U}(\mathcal{A}, \mathcal{D},$ $\mathcal{A}(\mathcal{D}), \mathcal{D}_{unlearn})$ = $\mathcal{A}(\mathcal{D}-\mathcal{D}_{unlearn})$. We call this method \emph{retrain from scratch}. Retrain from scratch is the only known exact unlearning method that is applicable to any learning algorithm. 

\myparatight{Limitations of retraining from scratch} A key limitation of retrain from scratch is that it incurs significant computation cost for unlearning complex classifiers such as deep neural networks. We aim to understand whether a pre-trained encoder can enable efficient retraining from scratch while  the unlearnt classifier still achieves high accuracy. 

We note that Cao and Yang~\cite{cao2015towards}  proposed exact unlearning methods for several simple learning algorithms such as support vector machines, naive Bayes, and decision trees, which are more efficient than retrain from scratch. Their methods are applicable to a learning algorithm that 1) relies on a summation form and 2) defines a unique globally optimal classifier for a given training dataset. However, these exact unlearning methods are not applicable to our three cases because the learning algorithms in our three cases do not satisfy the above two conditions.  Bourtoule et al.~\cite{bourtoule2021machine} showed that ensemble learning can be efficiently unlearnt. However, this method is also not applicable to our three cases since they are not ensemble learning. Therefore, we adopt retrain from scratch as the exact unlearning method in our experiments. 

\subsection{Experimental Setup}
\myparatight{Exact unlearning via retraining from scratch} Given a set of unlearning training examples $\mathcal{D}_{unlearn}$, we retrain a classifier from scratch  on the remaining training examples $\mathcal{D} - \mathcal{D}_{unlearn}$ in each case. Specifically, for {\caseI} and {\caseII}, we train classifiers on the remaining training examples from scratch. For {\LP}, we train the last fully connected layer on the remaining training examples from scratch. For {\FT}, we reset the encoder as its pre-trained parameters, re-initialize the last fully connected layer, and then train them on the remaining training examples. 

\myparatight{Evaluation metrics} We use \emph{testing accuracy} and \emph{retraining time} as evaluation metrics. Testing accuracy is the fraction of testing examples that are correctly classified by the unlearnt classifier. Retraining time  is the  time for training from scratch. 

\myparatight{Hyperparameter search and parameter setting}
Unless otherwise mentioned, we set the number of training examples to be unlearnt as 50. Note that we set the number of training examples to be unlearnt as 50 without the loss of generality as long as this number is small compared to the training data’s size. Moreover, we  sample $\mathcal{D}_{unlearn}$ uniformly at random  from a training dataset.  Given $\mathcal{D}_{unlearn}$, we use grid search to find the best combination of the number of training epochs and learning rate for retraining from scratch in each case, following the hyperparameter search process described in Section~\ref{sec:background}. In particular, the best hyperparameter setting of retraining from scratch achieves the highest accuracy on the validation examples. Since our $\mathcal{D}_{unlearn}$ are randomly sampled, we repeat each experiment five times and report the average results for each experiment. 

\subsection{Experimental Results}
\myparatight{Comparing {\LP} with {\FT}} Table~\ref{table:unlearn-case-3} shows the testing accuracy and retraining time of  unlearning for {\LP} and {\FT} on STL10 dataset. {\LP}  achieves a higher testing accuracy than {\FT} on STL10 dataset but lower testing accuracy than {\FT} on CIFAR10 and Tiny-ImageNet datasets (results are omitted for simplicity), which is consistent with our results in Table~\ref{comparison-non-adversarial}.  We observe that {\LP} is orders of magnitude more efficient than {\FT}. This is because {\LP} freezes the pre-trained encoder when training classifier, while {\FT} updates both the pre-trained encoder and the last fully connected layer. Therefore, we focus on {\LP} in the following experiments due to its computation efficiency. 

\begin{table}[!t]
\caption{Testing accuracy and retraining time for {\LP} and {\FT} when unlearning 50 training examples.}
\vspace{-5mm}
\label{table:unlearn-case-3}
\begin{center}
\begin{tabular}{|c|c|c|c|c|}
\hline
         & {Testing Accuracy}  &  {Retraining  Time (seconds)}  \\ \hline
{\LP}  &   0.984     &  $5.92 \times 10^1$     \\ \hline
{\FT}  &   0.962     &  $9.38 \times 10^3$         \\ \hline
\end{tabular}
\end{center}
\vspace{-5mm}
\end{table}

\begin{figure}[!t]
\vspace{-5mm}
    \centering
    \subfloat[Testing accuracy]{ 
    \label{unlearn-figure-testing-acc-fraction}
    \includegraphics[width=0.4\textwidth]{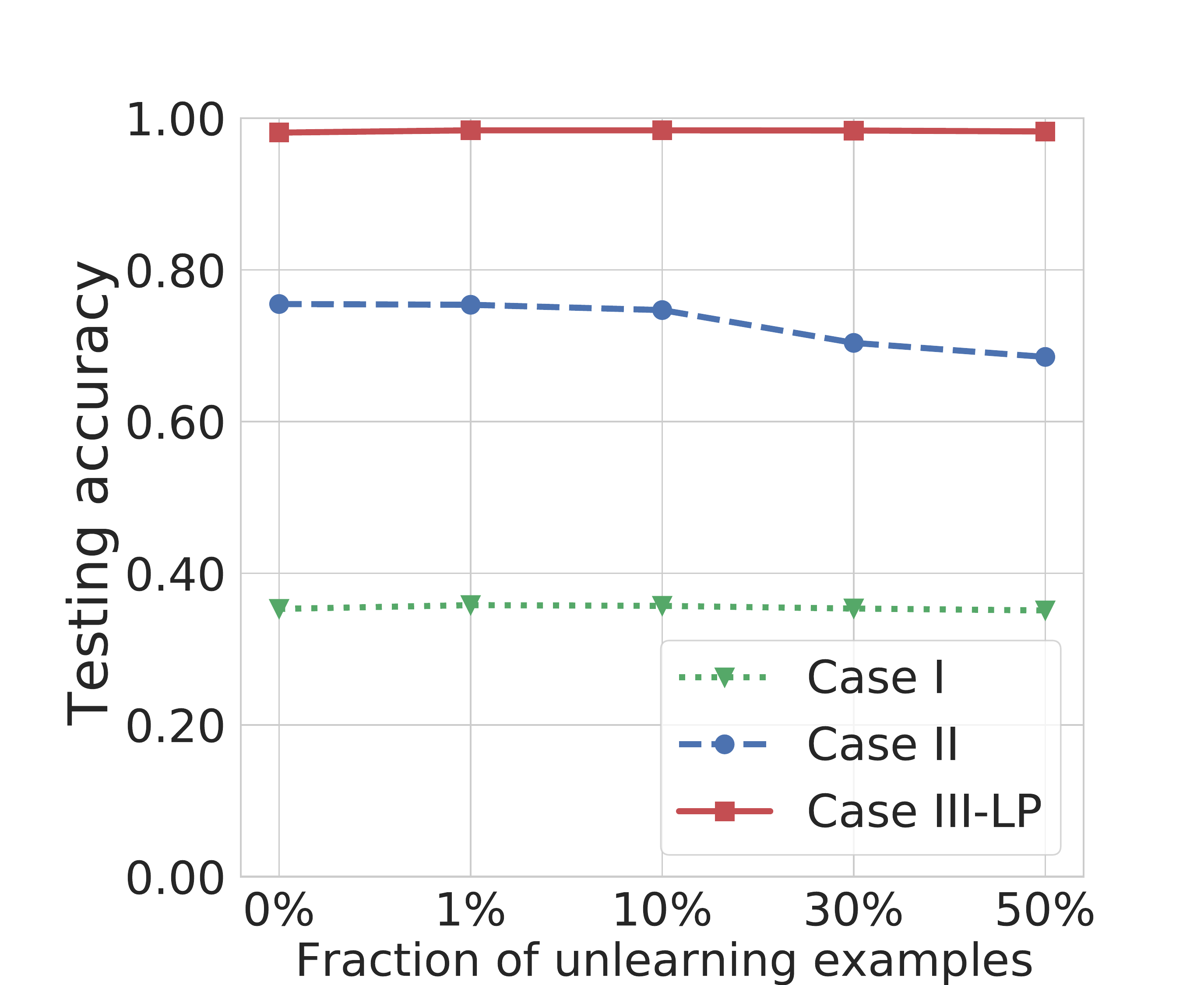}
    }
    \subfloat[Retraining time]{ 
    \label{unlearn-figure-cost-fraction}
    \includegraphics[width=0.4\textwidth]{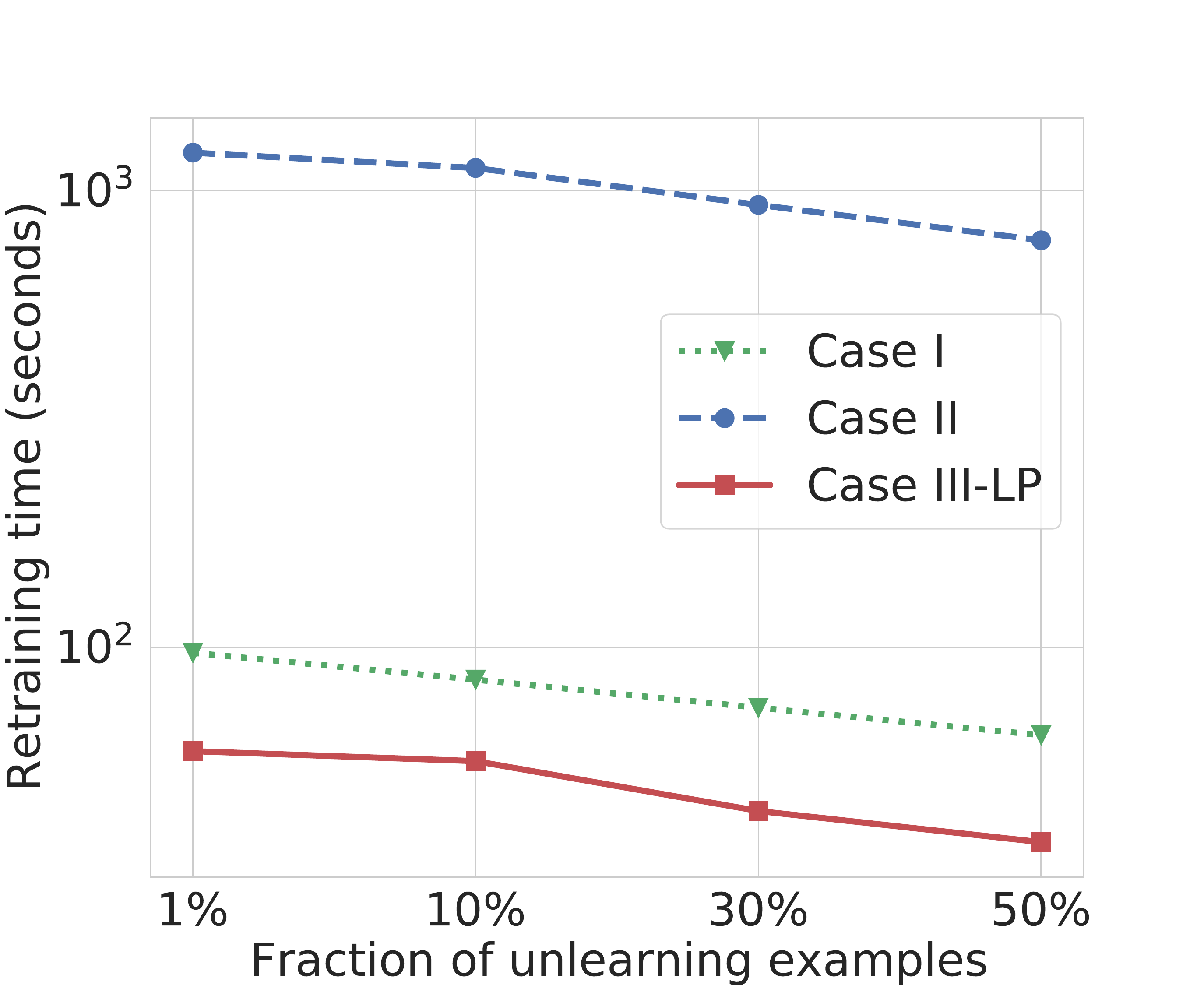}
    }
    \caption{Testing accuracy and retraining time for the three cases when unlearning different amount of training examples. } 
    \label{unlearn-figure-impact-fraction}
\end{figure}

\myparatight{Comparing the three cases}
Table~\ref{table:unlearningresults} shows the testing accuracy and retraining time of exact unlearning for the three cases when unlearning 50 training examples on STL10, while Figure~\ref{unlearn-figure-impact-fraction} shows the results when unlearning different amounts of  training examples. 
We observe that {\LP} achieves the highest testing accuracy among the three cases. This is because a pre-trained encoder extracts generalizable feature representations, which help improve testing accuracy.  Moreover, we observe that {\caseI} and {\LP} are orders of magnitude more efficient than  {\caseII}. This is because retraining the one fully connected layer in {\caseI} and {\LP} is much more efficient than retraining the deep neural network in {\caseII} from scratch. We also note that {\LP} is more efficient than {\caseI} on unlearning. The reason is that the fully connected layer in {\LP} has fewer parameters than the fully connected layer in {\caseI}. For example, on STL10, the fully connected layer in {\LP} has $1,024 \times 10 = 10,240$ parameters, where 1,024 is the dimension of the feature vector outputted by the encoder. However, the fully connected layer in {\caseI} has $96 \times 96 \times 3 \times 10 = 276, 480$ parameters, where $96 \times 96 \times 3$ is the dimension of training/testing inputs. Therefore, we conclude that unlearning in {\LP} is both much more accurate and efficient than {\caseII}, and is much more accurate and slightly more efficient than {\caseI}.

\begin{table}[!t]
\caption{Comparing the testing accuracy and retraining time of exact unlearning for the three cases when unlearning 50 training examples.}
\vspace{-5mm}
\begin{center}
\subfloat[Testing accuracy]{
\label{unlearn-results-compare-all-test-acc}
\begin{tabular}{|c|c|c|c|}
\hline
         &  STL10   & CIFAR10  & Tiny-ImageNet\\ \hline
Case I   &  0.358   &   0.247          &    0.072    \\ \hline
Case II &  0.754    & 0.946     &    0.506    \\ \hline
\LP     &  0.984    &   0.954          &   0.751    \\ \hline
\end{tabular}
}\\
\vspace{-2mm}
\subfloat[Retraining time (seconds)]{
\label{unlearn-results-compare-all-computation-cost}
\begin{tabular}{|c|c|c|c|}
\hline
         &  STL10   & CIFAR10   & Tiny-ImageNet\\ \hline
Case I   &  $  9.72\times 10^1 $    &   $  7.41\times 10^2 $          &    $  1.51\times 10^3 $   \\ \hline
Case II  &  $  1.21\times 10^3 $     &    $  8.63\times 10^3 $     &   $  1.82\times 10^4 $     \\ \hline
\LP &   $  5.92\times 10^1 $     &   $ 4.79\times 10^2 $       &   $  9.41\times 10^2 $    \\ \hline
\end{tabular}
}
\label{table:unlearningresults}
\end{center}
\vspace{-5mm}
\end{table}

\myparatight{Comparing LP with KNN in \caseIII} KNN classifies a testing input based on the labels of the $K$ nearest training inputs. Unlearning a KNN classifier is straightforward, i.e., we directly delete the training examples to be unlearnt from the training dataset. Therefore, we compare LP with KNN in {\caseIII} when a pre-trained encoder is available. Specifically, in KNN, we use the pre-trained encoder to compute a feature vector for each training/testing input; and when finding the $K$ nearest neighbors of a testing input, we  calculate the $\ell_1$ distance between a training input and the testing input based on their feature vectors.  

Table~\ref{unlearn-results-knn} shows testing accuracy and retraining time of unlearning LP and KNN in {\caseIII} in our default setting, as well as their inference time per testing input and storage after unlearning. The storage for LP is used to store the fully connected layer, while the storage for KNN is used to store the feature vectors and labels of the training inputs. 
We observe that KNN achieves comparable testing accuracy with LP and does not require any retraining time. However, KNN requires orders of magnitude more storage space and is orders of magnitude less efficient at inferring the label of a testing input. This is because KNN directly stores the training data and requires calculating distance between a testing input and every training input to predict its label. On the contrary, LP's storage and inference time for a testing input do not depend on the training dataset size. Therefore, as the training dataset size becomes larger, LP's advantages with respect to storage and inference time become more significant.    

\begin{table}[!t]
\caption{Comparing LP with KNN when unlearning 50 training examples.}
\vspace{-5mm}
\label{unlearn-results-knn}
\begin{center}
\begin{tabular}{|c|c|c|c|c|}
\hline
& \makecell{Testing \\accuracy}  &  \makecell{Retraining \\ time (seconds)} & \makecell{ Inference\\ time (seconds)} &\makecell{Storage \\ (MB)} \\ \hline
LP  &   0.984     &  $5.92 \times 10^1$     & $1.38 \times 10^{-5}$ & $2.20 \times 10^{-2}$   \\ \hline
KNN  &   0.978      &  0.00      & $3.1\times 10^{-3}$  &   $2.56 \times 10^{1}$  \\ \hline
\end{tabular}
\end{center}
\vspace{-5mm}
\end{table}

\myparatight{Summary} We observe that a pre-trained encoder improves the accuracy and/or efficiency of exact unlearning via retraining from scratch. Specifically, exact unlearning via retraining from scratch for a simple linear classifier in {\caseI} is efficient but not accurate. Retraining from scratch for a complex deep neural network in {\caseII} is more accurate than {\caseI}  but is not efficient.  Retraining from scratch for {\LP} is both accurate and efficient. 
\section{Conclusion, Limitations, and Future Work}
\label{sec:conclusion}
In this work, via a systematic and principled measurement study, we find that a high-quality encoder pre-trained on non-private data can improve the testing accuracy under no attacks, certified security guarantees, and/or efficiency of secure or privacy-preserving supervised (un)learning algorithms.  We acknowledge that our findings are limited to high-quality encoders pre-trained on non-private data, but state-of-the-art pre-trained encoders such as CLIP satisfy such conditions. One interesting future work is to understand whether and when our findings still hold for an encoder pre-trained on private data in adversarial settings.  
\bibliographystyle{plain}
\bibliography{bib}
\appendix

\begin{figure}[!h]
    \centering
    \subfloat[Testing accuracy]{ \label{poison-figure-impact-N-accuracy}
    \includegraphics[width=0.4\textwidth]{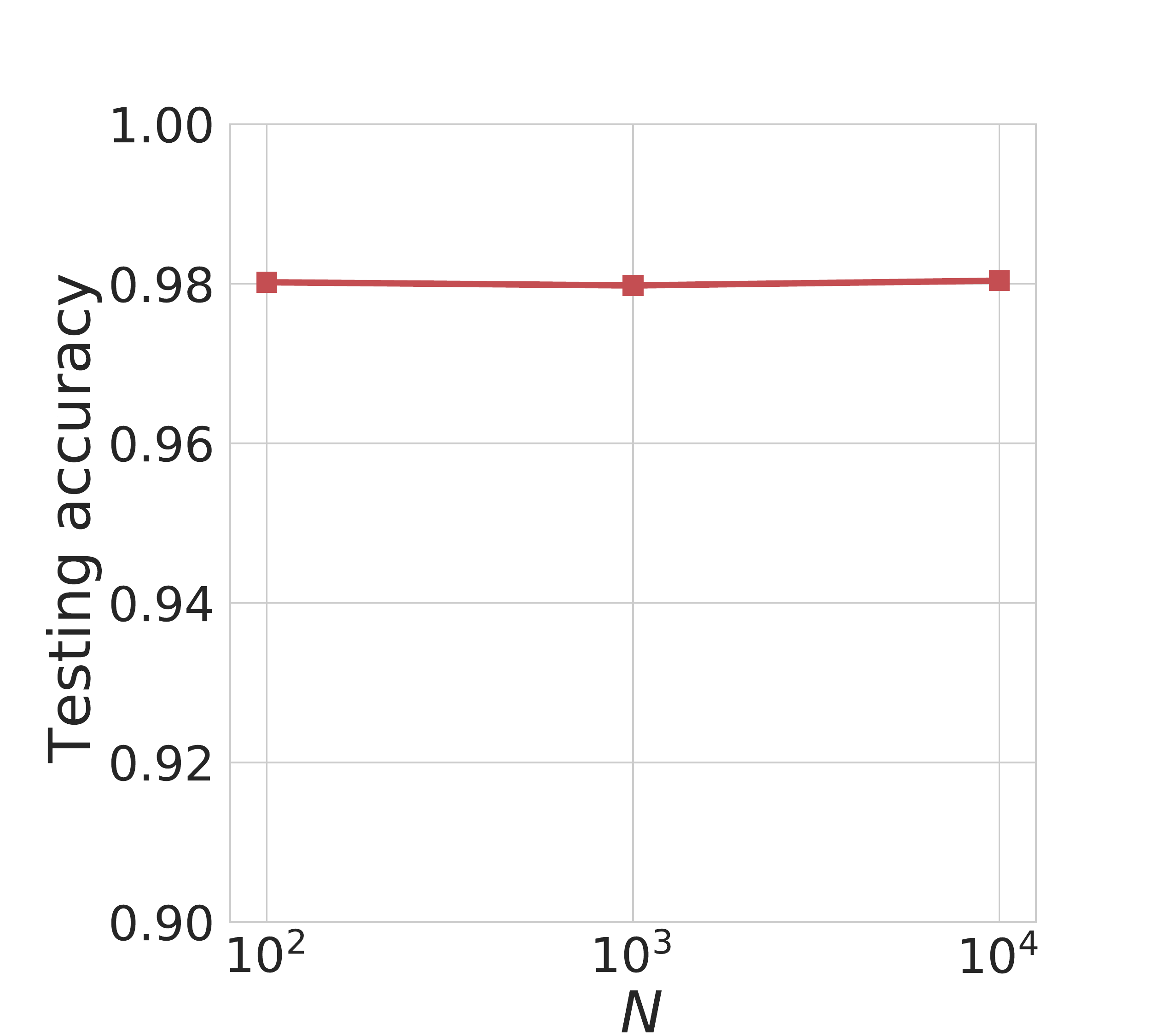}
    }
    \subfloat[ACPS]{
    \includegraphics[width=0.4\textwidth]{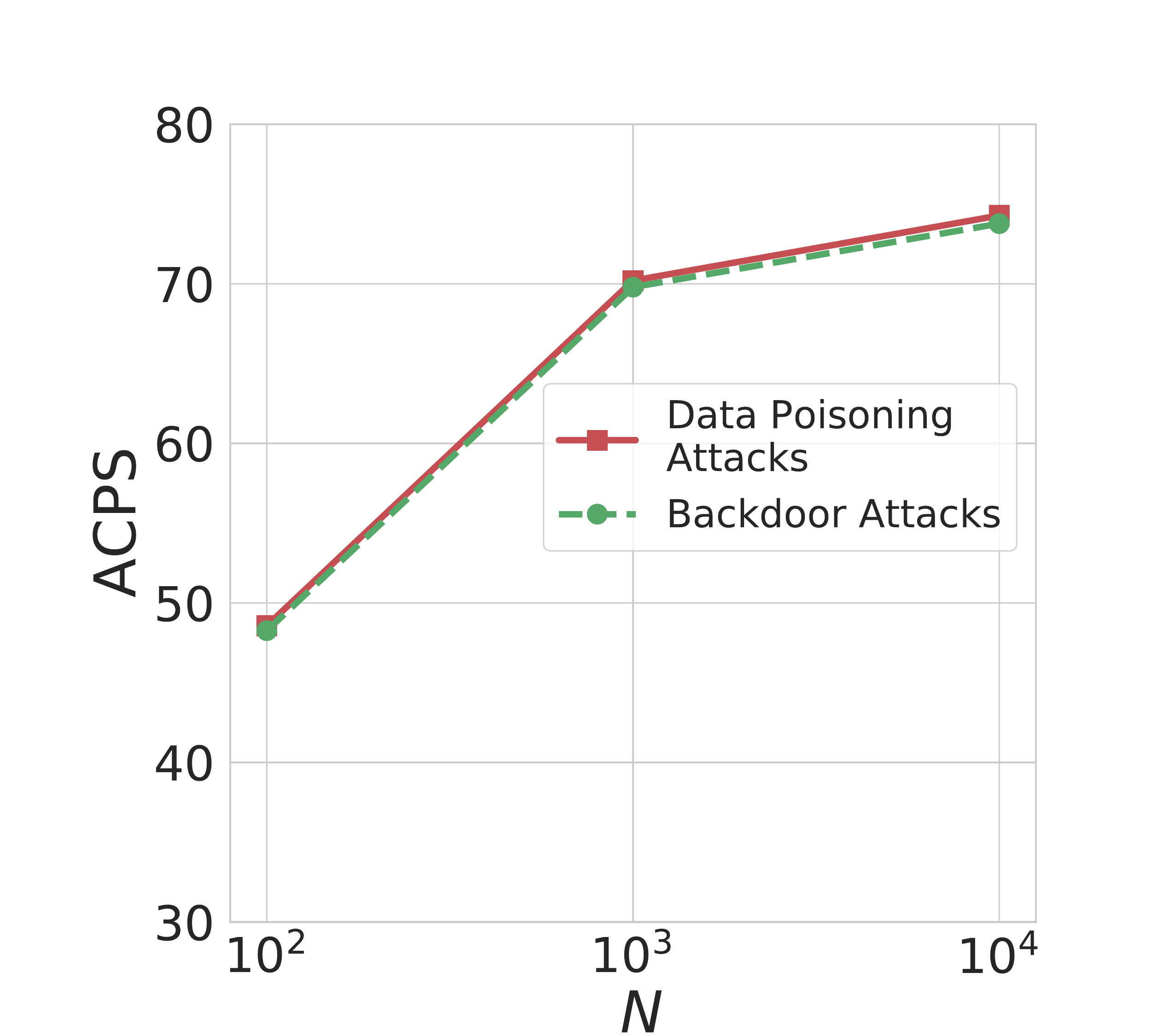}}
    \caption{Impact of $N$ on the testing accuracy of bagging under no attacks and ACPS of bagging against data poisoning and backdoor attacks.}
    \label{poison-figure-impact-N}
\end{figure}

\begin{figure}[!h]
    \centering
    \includegraphics[width=0.4\textwidth]{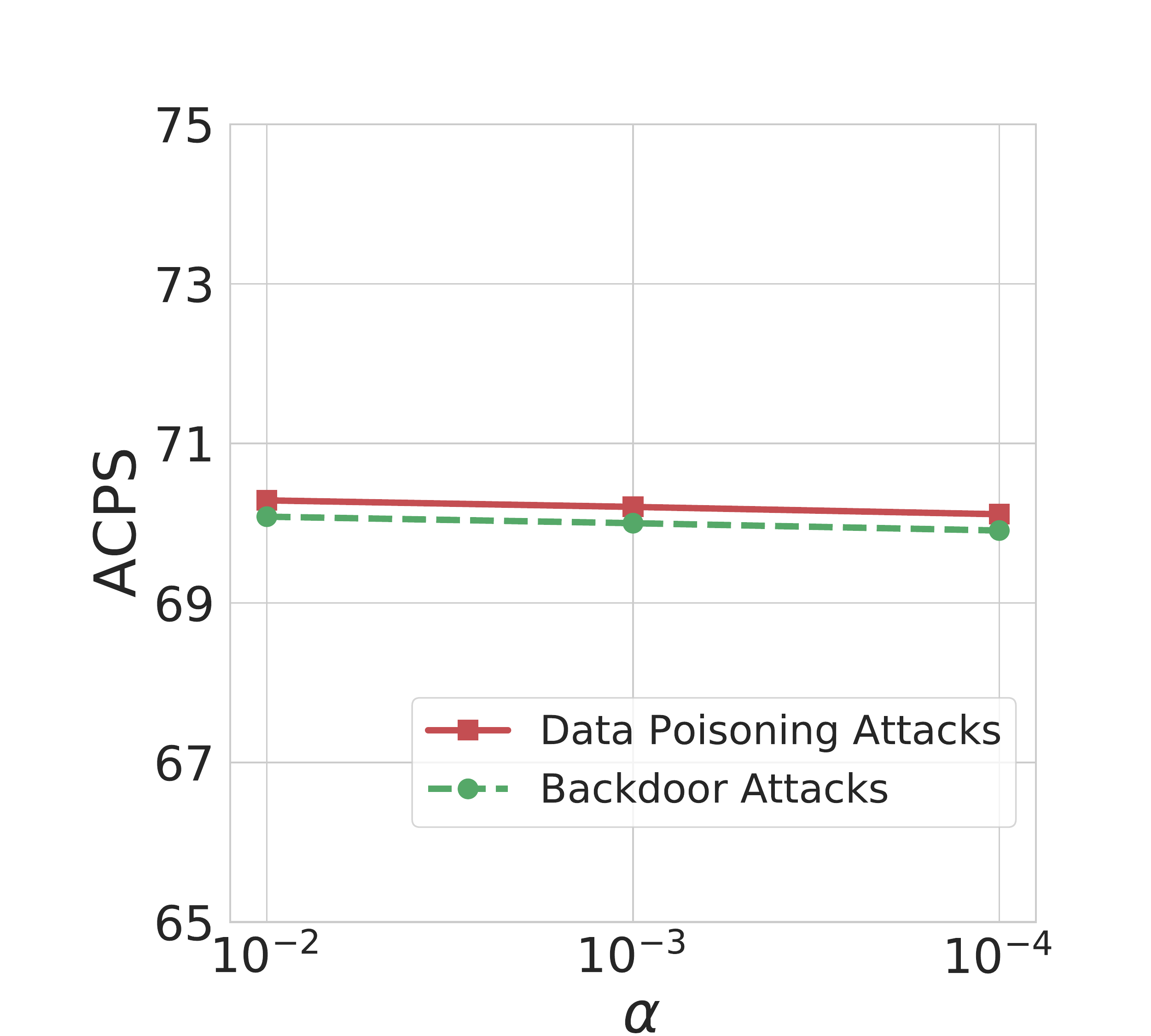}
    \vspace{2mm}
    \caption{Impact of $\alpha$ on the ACPS of bagging against data poisoning and backdoor attacks.}
    \label{poison-figure-impact-alpha}
\end{figure}

\begin{figure}[!h]
    \centering
    \subfloat[Testing accuracy]{\includegraphics[width=0.4\textwidth]{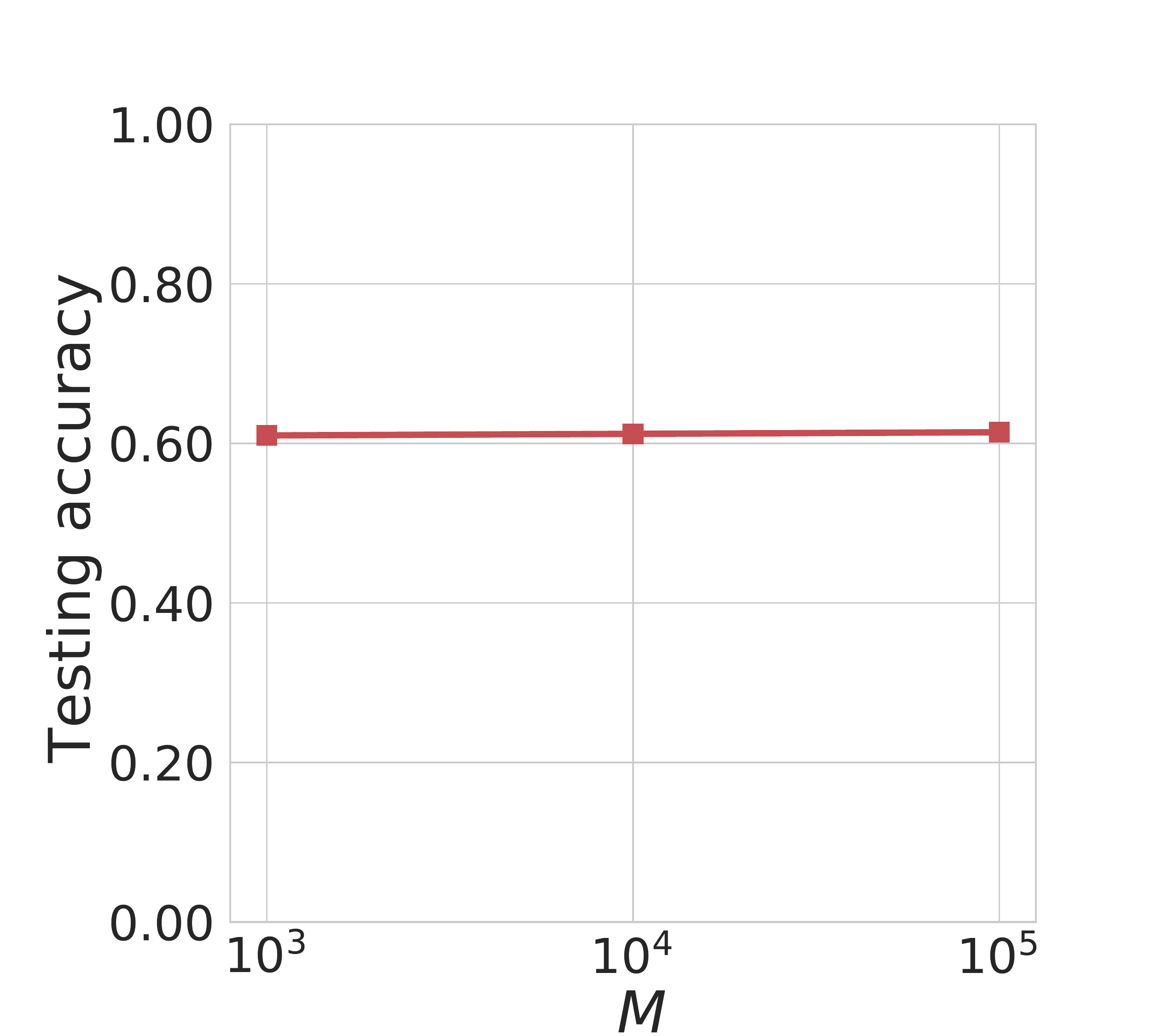}\label{adv-figure-impact-N-testing-accuracy}}
        \subfloat[ACR]{  \includegraphics[width=0.4\textwidth]{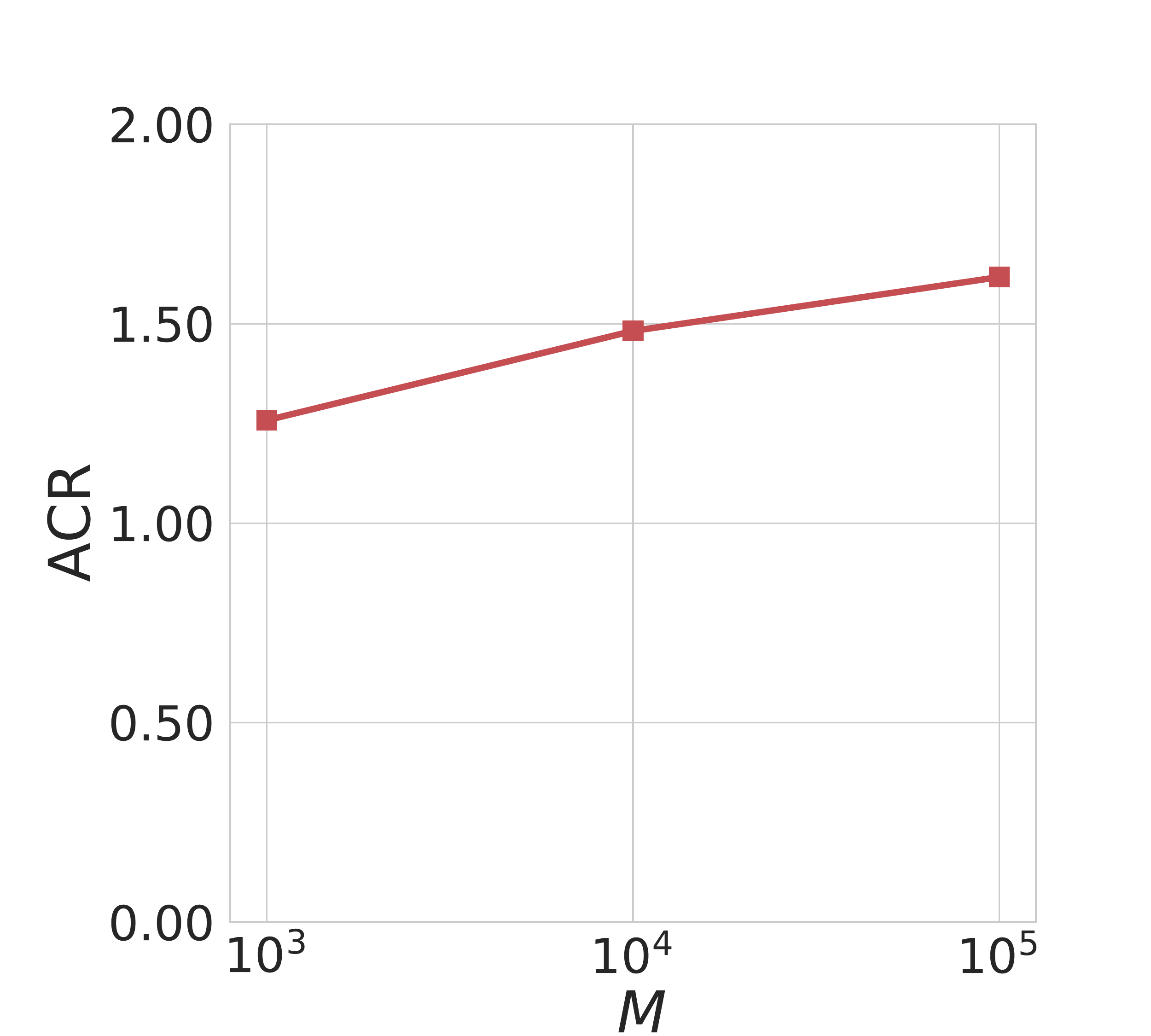} \label{adv-figure-impact-N-ACR} }
    \caption{Impact of $M$ on the testing accuracy and ACR of randomized smoothing.}
    \label{adv-figure-N-accuracy-ACR}
\end{figure}

\begin{figure}[!h]
    \centering
    \includegraphics[width=0.4\textwidth]{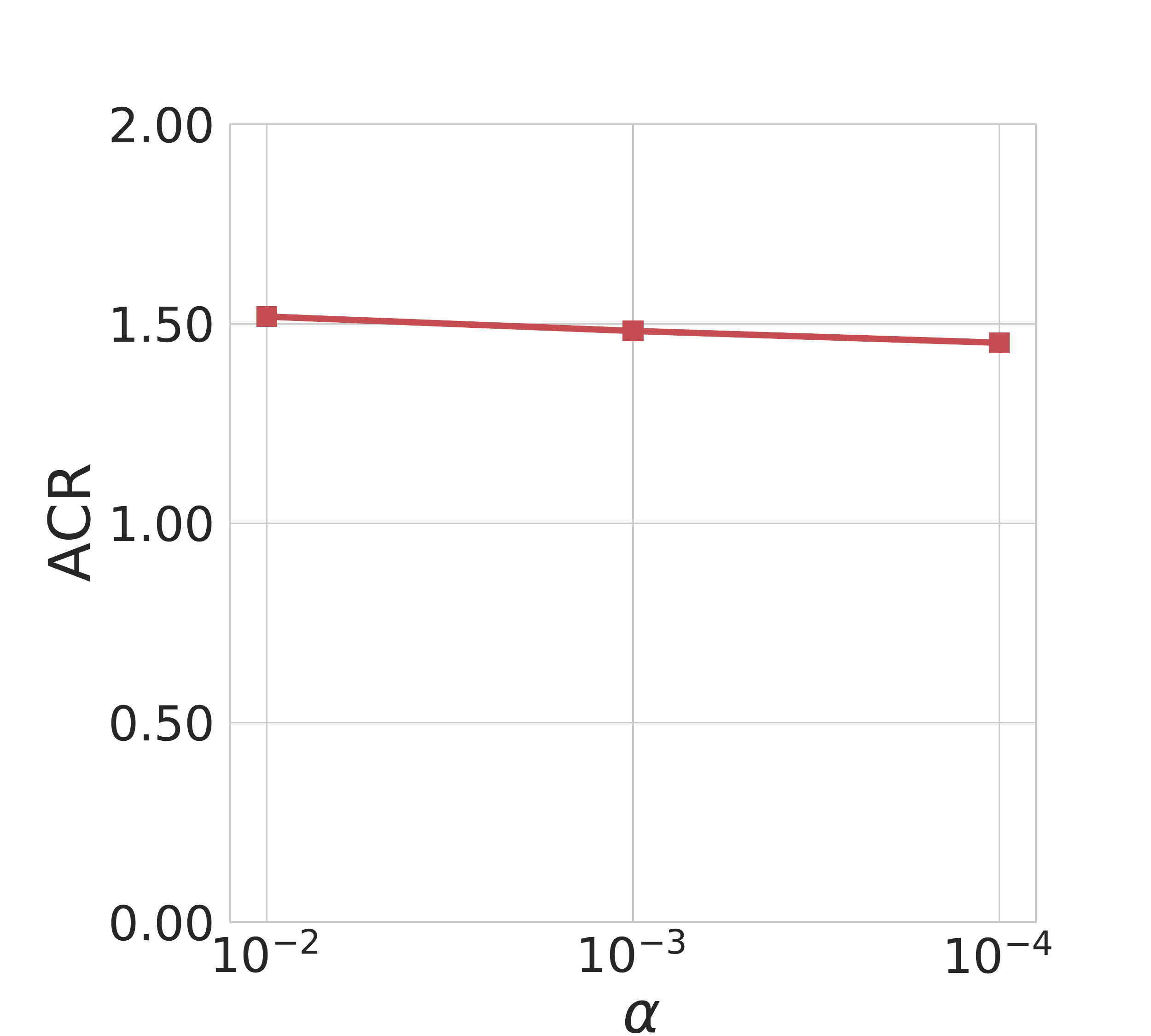} \label{adv-figure-impact-alpha} 

    \caption{Impact of $\alpha$ on the ACR of randomized smoothing.}
    \label{adv-figure-impact-alpha-ACR}
\end{figure}

\end{document}